\documentclass[10pt,a4paper]{article}
\usepackage{amsmath}
\usepackage{amsfonts}
\usepackage{amssymb}
\usepackage{cases}
\usepackage{dsfont}
\usepackage{color}
\usepackage{graphicx}
\usepackage{epsfig}
\usepackage{bm}
\usepackage{bbm}
\usepackage{ulem}
\usepackage{enumerate}

\newcommand\seq[3]{{#1}_{#2}^{#3}}
\newcommand{\bd}{\begin{document}}
\newcommand{\ed}{\end{document}}
\newcommand{\beq}{\begin{equation}}
\newcommand{\eeq}{\end{equation}}
\newcommand{\bef}{\begin{figure}}
\newcommand{\enf}{\end{figure}}
\newcommand{\bea}{\begin{eqnarray}}
\newcommand{\eea}{\end{eqnarray}}
\newcommand{\baR}{\begin{array}}
\newcommand{\eaR}{\end{array}}
\newcommand{\bc}{\begin{center}}
\newcommand{\ec}{\end{center}}
\newcommand{\ben}{\begin{enumerate}}
\newcommand{\een}{\end{enumerate}}
\newcommand{\bit}{\begin{itemize}}
\newcommand{\eit}{\end{itemize}}
\newcommand{\su}{\section}
\newcommand{\ssu}{\subsection}
\newcommand{\sssu}{\subsubsection}
\newcommand\ssssu[1]{\textbf{#1}}
\newcommand{\nid}{\noindent}
\newcommand{\nnb}{\nonumber}
\newcommand\bloc[2]{{\omega}_{#1}^{#2}}
\newcommand\blocp[2]{{\bra{\sigma \, \omega}}_{#1}^{#2}}
\newcommand{\un}{\mathbbm{1}}
\newcommand{\setZ}{\mathbbm{Z}}
\newcommand{\setN}{\mathbbm{N}}
\newcommand{\setR}{\mathbbm{R}}
\newcommand{\pare}[1]{\left(\, #1 \, \right)}
\newcommand{\scal}[1]{\langle\, #1 \, \rangle}
\newcommand{\bra}[1]{\left[\, #1 \, \right]}
\newcommand{\brac}[2]{\left[\, #1 \, | \, #2 \right]}
\newcommand{\Set}[1]{\left\{\, #1 \, \right\}}
\newcommand{\Setc}[2]{\left\{\, #1 \, ; \, #2 \right\}}
\newcommand{\Prob}[1]{\mathds{P}\left[\, #1 \, \right]}
\newcommand{\PT}[2]{\pi^{(T)}_{#2}\left[\, #1 \, \right]}
\newcommand{\pT}{\pi^{(T)}}
\newcommand{\Probex}[1]{P^{(\ast)}\left[\, #1 \, \right]}
\newcommand{\probc}[2]{\mathds{P}[#1 \, \left| \, #2 \right.]}
\newcommand{\Probc}[2]{\mathds{P}\bra{#1 \, \left| \, #2 \right.}}
\newcommand{\ProbC}[1]{\mathds{P}_\cC\bra{#1}}
\newcommand{\Pnc}[2]{mathds{P}_n\bra{#1 \, \left| \, #2 \right.}}
\newcommand{\Probcex}[2]{P^{(\ast)}\bra{#1 \, \left| \, #2 \right.}}
\newcommand{\Probcp}[2]{P^{(ap)}\bra{#1 \, \left| \, #2 \right.}}
\newcommand{\abs}[1]{\left|\, #1 \, \right|}
\newcommand{\deq}{\stackrel {\rm def}{=}}
\newcommand{\sqsubsetneq}{\stackrel {\tiny\sqsubset}{\tiny \neq}}
\newcommand\moy[1]{\mu\bra{#1}}
\newcommand\cA{{\cal A}}
\newcommand\cB{{\cal B}}
\newcommand\cC{{\cal C}}
\newcommand\cD{{\cal D}}
\renewcommand\H{{\cH_\bbeta}}
\renewcommand{\sb}{\sigma_B}
\newcommand\cF{{\cal F}}
\newcommand\cG{{\cal G}}
\newcommand\cI{{\cal I}}
\newcommand\cH{{\cal H}}
\newcommand\Q{\mathcal{Q}}
\newcommand\cK{{\cal K}}
\newcommand\cL{{\cal L}}
\newcommand\cM{{\cal M}}
\newcommand\cT{{\cal T}}
\newcommand\cS{{\cal S}}
\newcommand\cP{{\cal P}}
\newcommand\cO{{\cal O}}
\newcommand\cU{{\cal U}}
\newcommand\cV{{\cal V}}
\newcommand\cW{{\cal W}}
\newcommand\f{{\bm{f}}}
\newcommand\bF{{\bm{F}}}
\newcommand\bo{{\bm{o}}}
\newcommand\blambda{{\bm{\lambda}}}
\newcommand\bbeta{{\bm{\beta}}}
\newcommand\gk[1]{g_k\pare{#1}}
\newcommand\gLk{g_{L,k}}
\newcommand\akj[1]{\alpha_{kj}(#1)}
\newcommand\sif[1]{\seq{\omega}{-\infty}{#1}}
\newcommand\Xk[1]{X_k({#1})}
\newcommand\Ik[1]{I_k({#1})}
\newcommand\Skj[1]{S_{kj}({#1})}
\newcommand\XkD{\Xk{\bloc{0}{D-1}}}
\newcommand\Xkr{X_k^r}
\newcommand\tLk{\tau_{L,k}}
\newcommand\moyc[2]{\mu\bra{#1 \, | \, #2}}
\newcommand\Vkdet[1]{V_k^{(det)}({#1})}
\newcommand\Vksyn[1]{V_k^{(syn)}({#1})}
\newcommand\Vkext[1]{V_k^{(ext)}({#1})}
\newcommand\Vknoise[1]{V_k^{(noise)}({#1})}
\newcommand\sk[1]{\sigma_k({#1})}
\newcommand\skr{\sigma_k^r\pare{\bloc{0}{D-1}}}
\newcommand{\ent}[1]{\left[\, #1 \, \right]}
\newcommand\vm[1]{var_m\left[#1 \right]}
\newcommand\eg[1]{\stackrel {\rm #1}{=}}
\newcommand\tk[1]{\tau_k\pare{#1}}
\newcommand\tko{\tau_k(t,\omega)}
\newcommand\cBm[1]{\cB^{-1}\pare{#1}}
\newtheorem{theorem}{Theorem}
\newcommand{\bth}{\begin{theorem}}
\newcommand{\enth}{\end{theorem}}
\newcommand\omD[1]{\seq{\bra{\omega^{(#1)}}}{0}{D-1}}
\newcommand\omz[1]{\omega^{(#1)}(D)}
\newcommand\om[1]{\omega^{(#1)}}
\newcommand\Om[1]{\Omega^{(#1)}}
\newcommand\skjq{s_{kj;q}}
\newcommand\ikq{i_{k;q}}
\newcommand\skjqs{p_{kj;q_s}}
\newcommand\ikqs{i_{k;q_s}}
\newcommand\ikqsu{i_{k;q_{s_u}}}
\newcommand\ikqu{i_{k;q_u}}
\newcommand\xkq{x_{k;q}}
\newcommand\skjqv{s_{kj;q_{v}}}
\newcommand\tjr{t_j^{(r)}(\omega)}
\newcommand\dtt{\Phi(t,\omega)}
\newcommand\dts{\Phi(s,\omega)}
\newcommand\dto{e^{\Phi(t-t_0)}}
\newcommand\dtes{e^{\Phi(t-s)}}

\newcommand\rpf{R}
\newcommand\rpfc[1]{R\pare{#1}}
\newcommand\lpf{L}
\newcommand\lpfc[1]{L\pare{#1}}
\newcommand{\zb}{\zeta_{\bbeta}}
\newcommand{\etab}{\eta_{\bbeta}}
\newcommand{\mex}{\mu^{(\ast)}}
\newcommand{\phex}{\phi^{(\ast)}}
\newcommand{\Phex}{\Phi^{(\ast)}}
\newcommand{\Hex}{\cH_{\bbeta'}^{(\ast)}}
\newcommand\p[1]{\cP\bra{#1}}
\newcommand\w[2]{w_{#1}^{(#2)}}
\newcommand\Gb{\cG_\bbeta}
\newcommand\blp[3]{\omega^{#3,(#1)}_{#2}}
\newcommand\blpb[3]{\overline{\omega^{#3,(#1)}_{#2}}}
\newcommand\moyex[1]{\mu^{(\ast)}\bra{#1}}
\newcommand\E[1]{{\cal E}_{#1}}
\newcommand\Gap{\overline{G}}
\newcommand\gap[1]{\overline{g_{#1}}}
\newcommand{\K}[2]{\cK^{(#1)}(#2)}
\newcommand{\tO}[3]{\tau^{(#1)}(#2,#3)}
\newcommand{\Vr}{V_{reset}}
\newcommand{\bcoul}[2]{{\color{#1}{\textbf{#2}}}}
\newcommand{\bruno}[1]{\bcoul{red}{#1}}
\newcommand{\rod}[1]{\bcoul{blue}{#1}}
\newcommand{\tf}[2]{t_{#1}^{(#2)}}
\newcommand{\Tf}[1]{\mathcal{T}^{(#1)}(\omega)}
\newcommand{\Exp}[1]{\mathds{E}\left[\, #1 \, \right]}
\newcommand{\Expc}[2]{\mathds{E}\left[\, #1 \, | #2 \, \right]}
\newcommand{\Cov}[1]{Cov\left[\, #1 \, \right]}
\newcommand{\bD}[2]{\underline{D}_{#1}(#2,\omega)}
\newcommand{\hk}{\hat{k}}

\begin{document}
\title{Dynamics and spike trains statistics in conductance-based Integrate-and-Fire neural networks with chemical and electric synapses.}
\author{Rodrigo Cofr\'e, Bruno Cessac
\thanks{NeuroMathComp team (INRIA, ENS Paris, UNSA LJAD), Sophia Antipolis, France. \newline \indent INRIA, 2004 Route des Lucioles, 06902 Sophia-Antipolis, France. \newline \indent   email: bruno.cessac@inria.fr}}
\maketitle

\begin{abstract}
We investigate the effect of electric synapses (gap junctions) on collective neuronal dynamics and spike statistics in a conductance-based Integrate-and-Fire neural network, driven by a Brownian noise, where conductances depend upon spike history. We compute explicitly the time evolution operator and
show that, given the spike-history of the network and the membrane potentials at a given time, the further dynamical evolution can be written in a closed form. We show that spike train statistics is described by a Gibbs distribution whose potential can be approximated with an explicit formula, when the noise is weak. This potential form encompasses existing models for spike trains statistics analysis such as maximum entropy models or Generalized Linear Models (GLM). We also discuss the different types of correlations: those induced by a shared stimulus and those induced by neurons interactions.
\end{abstract}


\section{Introduction}

Communication between neurons involves chemical and electric synapses. Electric synapses transmission is mediated by gap junctions with a direct electrical communication between cells \cite{coombes-zachariou:07}, allowing faster communication than chemical synapses.
Electrical coupling between cells can be found in many parts of the nervous system \cite{bennett-zukin:04},\cite{connors-long:04}, and also outside: for example, certain cells in the heart and pancreas are connected by gap junctions \cite{keener-sneyd:98}.
At the network level, electric synapses have several prominent effects such as neurons synchronization \cite{beierlein-etal:00}, \cite{galarreta-hestin:99}, \cite{ostojic-etal:09}, and the generation of neural rhythms \cite{hormuzdi-etal:04}, \cite{bennett-zukin:04}. In the retina, gap junctions are involved in circuits containing the five  neuronal types (photo-receptors, horizontal, bipolar, amacrine and ganglion cells). In particular, amacrine and ganglion cells express numerous gap junctions, resulting in
extensive electrical networks of low-resistance pathways for direct electrical signaling between cells \cite{pan-paul-etal:10}. Unlike studies carried out in other parts of the Central Nervous System that often assign generic functions to electrical coupling
such as increased spike synchrony, studies in the retina have been able to detail the specific roles of individual gap junctions in visual processing.
Several distinct functional roles  have been elucidated for retinal gap junctions, related not only to the propagation of signals, but also to the encoding of specific visual information \cite{bloomfield:09}. In particular, electric synapses play a role in detection of approaching motion, and detection of differential motion \cite{gollisch-meister:10}.

Dealing with spike population coding,
it is clear that neurons interactions highly constrain the collective spike response of a neural assembly to stimuli. Therefore, unveiling the respective effect of chemical and electric synapses on spike responses is a mandatory step toward a better understanding of spike coding. As an example, the retina encodes visual stimuli in a highly parallel, adaptive and computationally efficient way, extracting specific stimulus features in a highly selective manner. Central to this ability is the segregation of retinal responses into a large number of parallel pathways, resulting in a mosaic of functional units encoding different aspects of the visual stimulus emerging from dedicated local retinal circuits (for a review see \cite{gollisch-meister:10}).
Those circuits involve both electric and chemical synapses. \\

On theoretical grounds, the role of gap junctions in shaping collective dynamics has been quite less studied than the role of chemical synapses, although 
different models and approaches have been used to address this problem, in the context of pattern formation, using techniques such as: Poincar\'{e} map \cite{chow-kopell:00,gao-holmes:07,pfeuty-mato-etal:05}, Lyapunov functions \cite{medvedev-wilson-etal:03}, mean-field approach and Fokker-Planck equation  \cite{ostojic-etal:09}, variance analysis \cite{medvedev:09}, and phase plane analysis \cite{coombes:08,coombes-zachariou:07}. The effects of gap junctions on spike trains statistics is even less known.

The goal of this paper is to push one step forward the mathematical analysis of the join effects of chemical and electric synapses on neurons dynamics and spike statistics. We propose and analyze a conductance-based Integrate-and-Fire model submitted to noise, inspired from the paper \cite{rudolph-destexhe:06}, with an extension including gap junctions. This 
work follows the paper \cite{cessac:11b} where only the effects of chemical synapses were studied. It was especially shown that spike statistics is
characterized by a Gibbs distribution whose potential can be explicitly computed.
This provided moreover a firm theoretical ground for recent studies attempting
to describe experimental rasters in the retina \cite{vasquez-marre-etal:12} as well as in the parietal
cat cortex \cite{marre-boustani-etal:09} by Gibbs distributions and maximum entropy principle. In the present paper, we
extend the mathematical analysis of \cite{cessac:11b} including electric synapses, allowing us to consider the simultaneous effects of both types of synapses in a single formalism. 

The main advantage of this type of model, rendering the mathematical analysis tractable, is that the sub-threshold dynamics of membrane potential is described by a linear system of stochastic differential equations (SDE). This system is nevertheless highly non trivial, as we show, since it is non autonomous, non homogeneous, with, additionally, a flow depending on the whole (spike) history via the chemical synapses. Moreover, electric synapses introduce another mechanism of history dependence, where the membrane potential of a neuron
depends on its past values, even those which are anterior to the last firing and reset of that neuron. The flow of this linear system of stochastic differential equations can be explicitly written. 

From this, one can compute a family of transition probabilities, characterizing the probability of a spike pattern at a given time, given the past. These transition probabilities define a Gibbs distribution characterizing spike train statistics. The potential of this Gibbs distribution can be approached by an explicit form, as we show. We have therefore an explicit characterization of spike statistics in a model including chemical and electric synapses.

This result has several implications in the realm of biological spike trains analysis. Especially, as we show, the Gibbs potential of our model encompasses existing models for spike trains statistics analysis such as maximum entropy models \cite{schneidman-berry-etal:06,tkacik-schneidman-etal:09,shlens-field-etal:06,shlens-field-etal:09,ohiorhenuan-mechler-etal:10,ganmor-segev-etal:11a,ganmor-segev-etal:11b,vasquez-marre-etal:12} and Generalized-Linear Models (GLM) \cite{pillow-paninski-etal:05,pillow-shlens-etal:08,pillow-ahmadian-etal:11,pillow-ahmadian-etal:11b,macke-cunningham-etal:11}. Moreover, our formalism affords the study non stationary dynamics, while stationarity is a major assumption when using maximum entropy models. Additionally, 
 as we show, gap junctions introduce major correlations effects ruining the hope of having conditionally-upon-the-past independent neurons, a central hypothesis in GLM. Finally, 
three types of effects are responsible for neuron correlations  (pairwise and higher order): shared stimulus, chemical couplings, and electric couplings. The two last types of correlations persists even when the stimulus is switched-off.\\

The paper is organized as follows. In section \ref{Sdef} we present the model, and detail how to handle properly the continuous-time dynamics of membrane potentials and the discrete-time dynamics of spikes. In section \ref{SMath}, we give explicit solutions for the system of equations presented in section \ref{Sdef}, where the network parameters as well as the network spike history dependence is explicit. In section \ref{Ssts} we describe the probability of spike events and the corresponding Gibbs distribution. The section \ref{SCons} is devoted to analyzing the possible consequences of those mathematical results on current neuroscience studies in the domain of MEA (Multi Electrodes Array) retina spike train analysis. Several conclusions and perspectives are drawn in the last section.

\section{Model definition} \label{Sdef}

The Integrate-and-Fire model remains one of the most ubiquitous model for simulating and analyzing the dynamics of neuronal circuits. Despite its simplified nature, it captures some of the essential features of neuronal dynamics (see \cite{lindner-schimansky:02,lindner-garcia:04,burkitt:06a,burkitt:06b,lindner:09} for a review). 
In this work we consider an extension of the conductance based Integrate-and-Fire neuron model proposed in \cite{rudolph-destexhe:06}. The model-definition follows the presentation given in \cite{cessac-vieville:08,cessac:11b}. Dynamics is ruled by a set of stochastic differential equations where parameters, corresponding to chemical conductances, depend upon a sequence of discrete variables summarizing the action potentials emitted in the past by the neurons. In this way, the dynamical system defined here is ruled both by continuous time and discrete time dynamical variables. Let us first introduce the discrete time variables.

\subsection{Spike trains}\label{SSpikeTrains}

We consider a network of $N$ neurons.
We  assume that there is a minimal time scale, $\delta$,  set to $1$
without loss of generality such that a neuron can at most emit an action potential  (\textit{spike}) within a time window of size $\delta$. This provides
a time discretization labeled with an integer time $n$. To each
neuron $k$ and discrete time $n$ one associates a spike variable
 $\omega_k(n)=1$ if  neuron $k$  as emitted a spike in the time window $[n\delta, (n+1)\delta[$ and $\omega_k(n)=0$
 otherwise. 
 The spike-state of
the entire network in time bin $n$ is thus described by a
  vector  $\omega(n) \deq \bra{\omega_k(n)}_{k=1}^{N}$, called a \textit{spiking pattern}.
A  \textit{spike block} is a finite ordered list of such vectors, written:
$$\bloc{n_1}{n_2} = \Set{\omega(n)}_{\{n_1 \leq n \leq n_2\}},$$
where spike times have been prescribed between time $n_1$ to $n_2$. 
A \textit{spike train} is a spike block $\bloc{-\infty}{+\infty}$. Obviously, real spike trains  start from some initial time $t_0 > -\infty$
and end at some final time $T< +\infty$, but, on mathematical grounds the consideration of bi-infinite sequences simplifies the analysis. To alleviate notations we simply write $\omega$ for a spike train.

The set of spiking pattern is denoted $\cA=\Set{0,1}^N$.
The set of  spike blocks $\bloc{m}{n}$ is denoted $\seq{\cA}{m}{n}$. The set of spike trains is denoted $\Omega=\cA^\setZ$. 


To each raster $\omega$ and each neuron index $j = \{1 \ldots N\}$ we associate an ordered (generically infinite) list
of ``firing times'' $\{\tf{j}{r}(\omega)\}_{r=1}^{+\infty} $ such that $\tf{j}{r} (\omega)$ is the $r$-th time of firing of neuron $j$ in the raster $\omega$.
$\tau_k(t,\omega)$ denote the last firing time of neuron $k$.

We note  $\ent{t}$  the largest integer which is $ \leq t$
(thus $\ent{-1.2}=-2$ and $\ent{1.2}=1$)
For a function $f : \setR \times \Omega \to \setR$, 
when we write $f(t,\omega)$ we mean $f(t,\bloc{-\infty}{\ent{t}})$. This corresponds to the physical notion of \textit{causality}. The functions $f(t,\omega)$ that we consider depend, at time $t$, upon spikes occurring before $t$.

\subsection{Membrane potential dynamics}

Neurons are considered here as points, with neither spatial extension nor biophysical structure (axon, soma, dendrites). 
We note $V_k(t)$ the membrane potential of neuron $k=1, \dots, N$, at time $t$. Denote $V(t)$ the vector with entries $V_k(t)$. The continuous-time dynamics of $V(t)$ is defined as follows.
Fix a real variable $\theta>0$ called ``firing threshold". 
For a fixed time $t$, we have two possibilities:

\begin{enumerate}

\item Either $V_k(t) < \theta$, $\forall k=1, \dots,N$. This corresponds to \textit{sub-threshold dynamics}.

\item Or, $\exists k$, $V_k(t) \geq \theta$. Then, we speak of \textit{firing dynamics}.

\end{enumerate}

\sssu{Sub-threshold dynamics}

The sub-threshold variation of the membrane potential of neuron $k$ at time $t$ is given by:
\begin{equation}\label{sub_threshold}
C_k \frac{dV_k}{dt} = -g_{L,k}(V_k-E_{L})-\sum_j g_{kj}(t,\omega)(V_k-E_j) - \sum_j \gap{kj} \left( V_k - V_j\right)+I_k(t).
\end{equation}
\begin{itemize}
\item $C_k$ is the membrane capacity of neuron $k$.

\item The term $I_k (t)$ is a current given by:
\begin{equation}\label{current}
I_k (t) = i_k^{(ext)}(t) + \sigma_B \xi_k (t),
\end{equation}
where $i_k^{(ext)}(t)$ is a deterministic external current (``stimulus").
The noise term  $\xi_k (t)$ mimics: (i) the random variation in the ionic flux of
charges crossing the membrane per unit time at the post synaptic button, upon opening of ionic channels
due to the binding of neurotransmitter, (ii) the fluctuations in the current resulting from the large number of opening and closing of ion channels \cite{schwalger-etal:10,linaro-etal:11}; (iii) noise coming from electrical synapses. It is common to model this noise by a Wiener white noise (diffusion approximation). We use this modelling choice in the paper.
We note $\sigma_B > 0$ the amplitude (mean-square deviation) of 
$\xi_k (t)$.

\item $g_{L,k}$ is a leak conductance, $E_{L} < 0$ is the leak reversal potential.

\item  $g_{kj}(t,\omega)$ corresponds to the conductance of the chemical synapses from the pre-synaptic neuron $j$ to the post-synaptic neuron $k$, while $E_j$ is the reversal potential associated with the neurotransmitter emitted by neuron $j$. In this model, the conductance $g_{kj}(t,\omega)$ at time $t$ depends upon the spikes emitted by the pre-synaptic neuron $j$. The description of this dependence is made explicit in section \ref{Sconductances}.
 
\item $\gap{kj}$ is the electric conductance (gap junctions) between two different neurons $j$ and $k$. Although, the gap junction strength is biophysically non static \cite{pan-paul-etal:10,hu-pan-etal:10}, we model it here  as  a simple ohmic conductance, tending to equalize membrane potentials of the neurons they connect. 
As a consequence, $\gap{kj}=\gap{jk} \geq 0$.

\item For chemical as well as electric contacts terms of eq. (\ref{sub_threshold}) the sums $\sum_j$ hold on all neurons, but we afford some conductances $g_{kj}(t,\omega)$ or $\gap{kj}$ to be $0$, corresponding to no connection (chemical or electrical), between neuron $j$ and neuron $k$. In this way, we can define  chemical and electric network topologies. The mathematical results obtained in section \ref{SMath} hold for any such network.
\end{itemize}

\sssu{Update of chemical synapses conductances}\label{Sconductances}

Upon firing of the pre-synaptic
neuron $j$ at (discrete) time $\tf{j}{r} (\omega)$, the membrane conductance $g_{kj}(t)$ of the post-synaptic neuron $k$ is modified as \cite{rudolph-destexhe:06}:
\begin{equation}\label{gkj_one_spike}
g_{kj}(t)=g_{kj}(\tf{j}{r} (\omega))+ G_{kj}\alpha_{kj}(t-\tf{j}{r} (\omega)), \quad t>\tf{j}{r} (\omega),
\end{equation}
where $G_{kj} \geq 0$ characterizes the maximal amplitude of the conductance during a
post-synaptic potential. 

The function $\alpha_{kj}$ (called ``alpha'' profile \cite{destexhe-mainen-etal:98}) mimics the time course of the chemical synaptic conductance upon the occurrence of the spike.
We assume that the alpha profiles have the form:
\beq\label{alpha}
\alpha_{kj}(t) = h(t)e^{-\frac{t}{\tau_{kj}}}H(t),
\eeq
with $\alpha(0) =0,$ where $h(t)$ is a polynomial function continuous at 0, typically $h(t)=\frac{t}{\tau_{kj}},$ where $\tau_{kj}$ is the characteristic decay time. $H(t)$ is the Heaviside function.

As a consequence, upon the arrival of spikes at times $\tf{j}{r} (\omega)$ in the time interval $\left[ s, t\right]$, eq. (\ref{gkj_one_spike}) becomes:
$$g_{kj}(t)=g_{kj}(s)+G_{kj} \sum_{r;s \leq \tf{j}{r}(\omega) <t}\alpha_{kj} \left( t-\tf{j}{r}(\omega)\right).
$$

If we assume that this relation extends to $s \to -\infty$ and if we set $\lim_{s \to -\infty}g_{kj}(s)=0$
(see \cite{cessac:11b} for a justification), we finally obtain:
\beq\label{gkj}
g_{kj}(t,\omega)=G_{kj} \sum_{r; \tf{j}{r}(\omega) <t} \alpha_{kj} \left( t-\tf{j}{r}(\omega)\right), \nonumber
\eeq
The notation $(t,\omega)$ makes explicit the dependence of the conductance upon the (past) firing activity of the pre-synaptic neuron $j$. This dependence decays exponentially fast thanks to the exponential decay of $\alpha_{kj}$.

\subsubsection{Firing dynamics}

If, at time $t$, some neuron $k$ reaches its threshold $\theta$, $V_k(t)=\theta$, then this neuron elicits a spike. In neurophysiology, the spike emission is a complex nonlinear mechanism involving additional physical variables (probability of opening/closing of ionic channels), as described by the Hodgkin-Huxley equations \cite{hodgkin-huxley:52}. The time course of the spike has a typical shape (depolarization / repolarization / refractory period) with a finite duration. One important simplification in Integrate-and-Fire models is to describe neuron dynamics in terms of the membrane potential only. As a consequence, the spike shape has to be simplified.
In the simplest Integrate-and-Fire models \cite{gerstner-kistler:02b,ermentrout-terman:10} a spike is registered at time $t$ whenever  $V_k(t)=\theta$, whereas the membrane potential is instantaneously reset to $0$. Moreover, the refractory period and transmission propagation delays are set to $0$. Beyond the bio-physical fact that reset, refractory period and transmission propagation are not instantaneous, this modelling leads to severe mathematical problems and logical inconsistencies, such as the possibility of having uncountably many spikes within a finite time interval, or situations where the state of a neuron cannot be defined (see \cite{cessac-vieville:08, cessac:10a} for a discussion).

To avoid those problems we consider that spikes emitted by a given neuron are separated by a minimal time scale
$\tau_{sep}>0$. Additionally, to conciliate the continuous time dynamics of membrane potentials and the discrete time dynamics of spikes we define the spike and reset as follows.

\begin{enumerate}

\item The neuron membrane potential $V_k$ is reset to $0$
at the \textit{next integer time} (in $\delta$ units) after $t$, namely $[t]+1$. Between $t$ and $[t]+1$ the membrane potential keeps on evolving according to (\ref{sub_threshold}). The main reason for this modelling choice is that it makes the mathematical analysis simpler. 

\item A spike is registered at time $[t]+1$. This allows us to represent spike trains as events on a discrete time grid. It has the drawback of artificially synchronizing spikes coming from different neurons, in the deterministic case 
\cite{cessac-vieville:08,kirst-timme:09}. However, the presence of noise in membrane potential dynamics eliminates this synchronization effect. 

\item We consider that $\tau_{sep}>0$ is a multiple of $\delta$ (thus an integer). 

\item Between $[t]+1$ and $[t]+\tau_{sep}$ the membrane potential $V_k$ is maintained to $0$ (refractory period).
From time $[t]+\tau_{sep}$ on, $V_k$ evolves according to (\ref{sub_threshold}) until the next spike.

\item When the spike occurs (at time $[t]+1$), the raster $\omega$ as well conductances $g_{kj}(t,\omega)$ are updated. 
 
\end{enumerate}

\begin{figure}[htbp]
\begin{center}
\includegraphics[height=5cm,width=7cm]{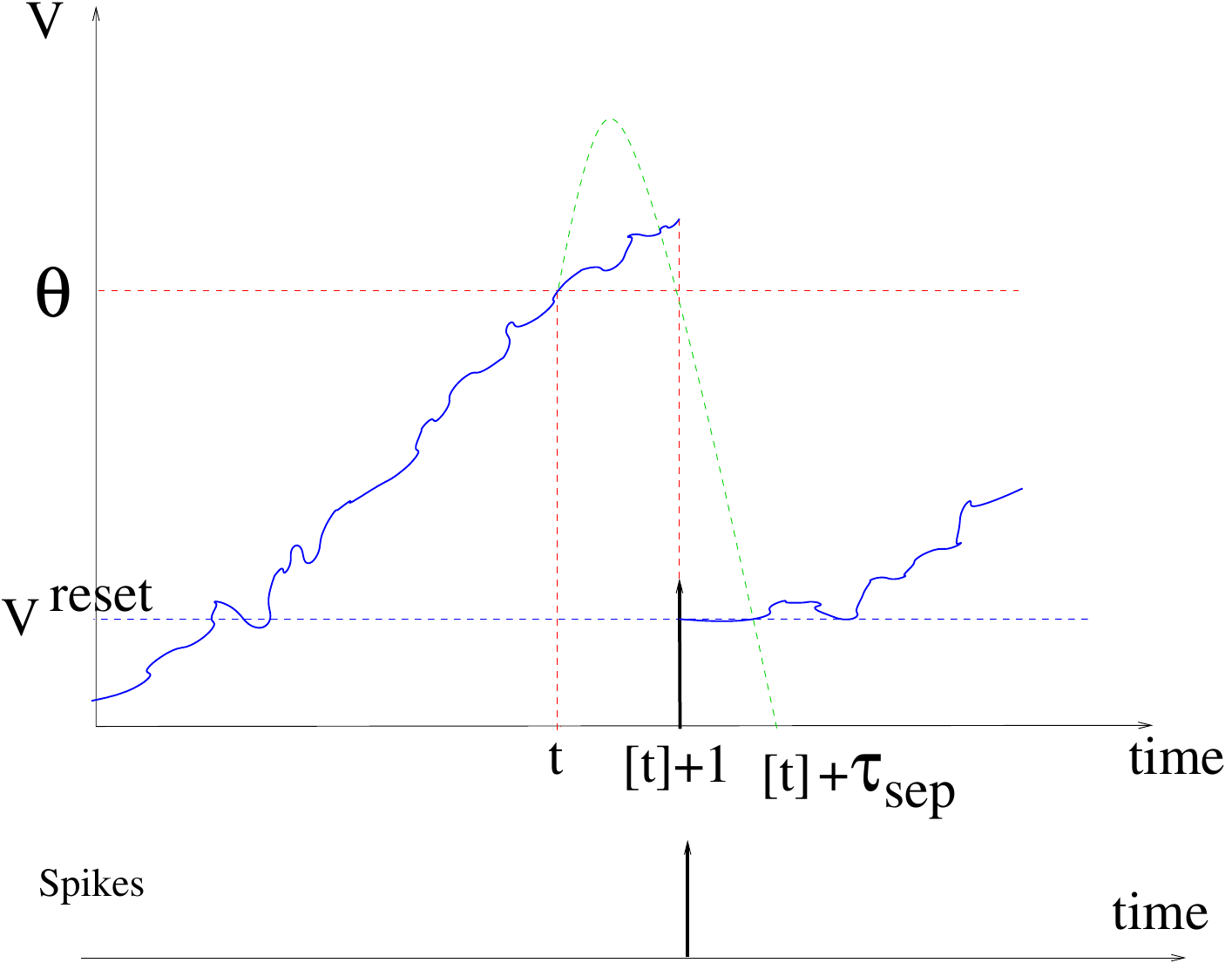} 
\caption{Top: Continuous time course of the membrane potential in our model. The green dashed curve illustrates the shape
of a real spike, but what we model is the blue curve. Bottom: A spike is registered at integer time $[t]+1$.}
\label{F1gap}
\end{center}
\end{figure}

Figure \ref{F1gap} illustrates these modelling choices.

\sssu{Matrix-Vector representation of subthreshold dynamics}

The sub-threshold dynamics can be rewritten in the form of a stochastic linear non-autonomous and non-homogeneous differential equation:
\begin{equation}\label{sub_threshold_vec}
C \frac{dV}{dt}+ \bra{G(t,\omega) - \Gap} \, V=I(t,\omega),
\end{equation}
where $C$ is a diagonal matrix which contains the capacity of each neuron. For simplicity we assume from now on that all neurons have the same capacity $c$ so that $C=c \cI_N$ where $\cI_N$ is the identity matrix of dimension $N$; $V$ is the vector of membrane potentials; $G(t,\omega)$ is a diagonal matrix:
\begin{equation}\label{gktomega}
G_{kl}(t,\omega)=\bra{
g_{L,k} 
\, + \, \sum_{j=1}^N g_{kj}(t,\omega)
} \, \delta_{kl} \deq g_k(t,\omega) \delta_{kl}. 
\end{equation}
$\Gap$ is a matrix with entries $\gap{kj}$ for $k\neq j$ and $-\sum_{j}\gap{kj}$ for $k=j$. It is therefore symmetric and the non diagonal part of this matrix specifies the connection topology of electric synapses in the network.
Finally, $I(t,\omega)$ is the vector of currents that can be separated in 3 components.
\begin{equation}\label{I(t)}
I(t,\omega)= I^{(cs)}(t,\omega) +I^{(ext)}(t)+ I^{(B)}(t),
\end{equation}
with:
\beq\label{Ics}
I_k^{(cs)}(t,\omega)= \sum_j W_{kj} \alpha_{kj}(t,\omega),
\eeq
where:
\beq\label{Wkj}
W_{kj} \deq G_{kj} E_j,
\eeq
is the synaptic weight from neuron $k$ to neuron $j$;
\beq\label{Iext}
I_k^{(ext)}(t) = g_{L,k}E_L + i_k^{(ext)}(t);
\eeq
and
$$
I_k^{(B)}(t)= \sigma_B \xi_k(t).
$$
The first term corresponds to the current received by neuron $k$ from chemical synapses (cs), the second term represents the external current (ext) and the last term (B) is the the noise part of the current.\\

Define:
\beq\label{Phi}
\dtt= C^{-1}\left( \Gap - G(t,\omega)\right),  \eeq
 which is a symmetric matrix. All entries of $\dtt$ are bounded and continuous in time (alpha profiles are continuous in time and the components of $\dtt$ are composition of continuous functions). \\

Defining:
\beq\label{f}
f(t,\omega)= C^{-1}I^{(cs)}(t,\omega)+C^{-1}I^{(ext)}(t),  
\eeq 
and using the decomposition of currents (\ref{I(t)}) the system (\ref{sub_threshold_vec}) can be expressed as a system of coupled Stochastic Differential Equations (SDE) of Ornstein-Uhlembeck type (O-U) in $\setR^N$
$$
\frac{dV}{dt} = \underbrace{C^{-1}(\Gap - G(t,\omega))}_{\dtt}V +\underbrace{C^{-1}I^{(cs)}(t,\omega)+C^{-1}I^{(ext)}(t)}_{f(t,\omega)} + C^{-1}I^{(B)}(t),
$$
$$\Leftrightarrow$$
\begin{equation}\label{DNN}
\left\{
\begin{array}{cclc}
dV & = & (\dtt V +f(t,\omega))dt + \frac{\sigma_B}{c} \cI_N dW(t), \\
&&\\
V(t_0) & = & v,
\end{array}
\right.
\end{equation}

\nid where $v$ is the initial condition at time $t_0$.
Here $f(t,\omega)$ is thus a non random, measurable and locally bounded function of $t$; $\frac{\sigma_B}{c}$ is a constant; and $W(t)$ is a standard $N$-dimensional Brownian motion independent of $v \in \setR^N$.

\sssu{Remarks}\label{SRemarksFrame}

\begin{itemize}

\item Although (\ref{DNN}) is a linear system, it has a rather complex structure, due to the $\omega$-dependence of $\Phi(t,\omega)$, $f(t,\omega)$. Indeed, these functions integrate the past spike-activity of the network from the initial time $t_0$ to time $t$. 
Thus, the membrane potentials at time $t$ are determined by the past spikes-activity which, in turn, is determined by the trajectory $V(s), s \in [t_0, t]$ of the membrane potentials: the evolution of the network depends on its whole history via $\omega$. If $\omega$ is given, the integration of (\ref{DNN}) generates a flow which is explicitly computed in section \ref{S_exist_uniq}.

\item  The discrete structure of the spike trains set $\Omega$ (discrete events and discrete times) induces a partition on the set $\mathcal{P}$ of trajectories of $V$. A trajectory  
$\cV \in \mathcal{P}$ belongs to the partition element $\mathcal{P}_\omega$, associated with the spike train $\omega$, if and only if:
\beq\label{Compatibility}
\forall k=1, \dots N, \, \forall n \in \setZ, \quad
\left\{
\begin{array}{ccc}
\omega_k(n)=0 \, \Leftrightarrow\, \max_{t \in ]n-1,n]} V_k(t)<\theta;\\
\\
 \omega_k(n)=1 \, \Leftrightarrow\, \max_{t \in ]n-1,n]}
 V_k(t) \geq \theta;
\end{array}
\right.
\eeq
This constitutes \textit{compatibility conditions} between spike trains and membrane potential trajectories. 

In the absence of noise ($\sigma_B=0$) some partition elements $\mathcal{P}_\omega$ (depending on model-parameters) are not visited by any trajectory.
It has been shown in some variants of (\ref{DNN}) considered in \cite{cessac:08,cessac-vieville:08}, 
that the set of non empty $\mathcal{P}_\omega$s is finite, leading to specific, although quite rich, periodic orbit structure of the attractors. In the presence of noise, 
all $\mathcal{P}_\omega$s are visited by any trajectory with a positive probability. 

\item The Wiener process on noise trajectories induces a probability measure $\mu$ on rasters, described in section \ref{Ssts}, characterizing spike train statistics. 

 \end{itemize}
 
\subsection{Examples}

The model (\ref{sub_threshold}) is quite general as it considers chemical and electric synapses, with a spike-history dependence. We don't know any analysis of this model in its most general form in the literature. However, upon simplifications it reduces to several models studied in the literature. 

\subsubsection{No electric synapse}

As mentioned in the introduction the model without  electric synapse has been studied in  \cite{cessac:11b}. The sub-threshold dynamics reads: 
%
$$
C_k \frac{dV_k}{dt} = -g_{L,k}(V_k-E_{L})-\sum_j g_{kj}(t,\omega)(V_k-E_j)+I_k(t).
$$
%
In this case 
$$\Phi(t,\omega)=C^{-1}G(t,\omega),$$ 
while $f(t,\omega)$ takes the same form (\ref{f}) as in the general model.

\subsubsection{Simple chemical conductances}

By ``simple" we mean a conductance matrix that takes the form $G(t,\omega) \equiv 
\kappa(t,\omega)\mathcal{I}_N$, where $\kappa(t,\omega)$ is a real function. Therefore, at time $t$ and given an history $\omega$, all neurons have the same conductance $\kappa(t,\omega)$, and the sub-threshold equation (\ref{sub_threshold_vec}) reads:
\begin{equation}\label{ssc}
C \frac{dV}{dt} = -\kappa(t,\omega) \, \mathcal{I}_N \, V  + \Gap V + I^{(cs)}(t,\omega) + I^{(ext)}(t) + \sigma_B  dW(t).
\end{equation}

On mathematical grounds being ``simple" is a general condition ensuring that $G(t)$
and $\Gap$ commute (see section \ref{SExpFlow}). On practical grounds, classical examples are $\kappa(t,\omega)=0$
or $\kappa(t,\omega)=cste$. 

Prominent illustrations are the Medvedev model \cite{medvedev:09} and the Ostojic-Brunel-Hakim model \cite{ostojic-etal:09}.

\section{Solutions of the stochastic differential equation}\label{SMath}

We now derive several mathematical results allowing the integration of the SDE (\ref{DNN}) and the consideration of firing dynamics.

\ssu{Flow in the sub-threshold regime}

We consider first the integration of (\ref{DNN}) on a time interval $[t_0,t]$ in the sub-threshold regime, $V_k(s) < \theta$, $k=1, \dots, N$, $s  \in [t_0,t]$. We assume that $\omega$ is fixed. Note that, necessary, $\omega$ obeys $\omega_k(n)=0$, $k=1,\dots, N$, $t_0 < n <t$ (cf the compatibility conditions (\ref{Compatibility})). Nevertheless, we don't make any assumption on $\omega$ before time $t_0$, that is, we can have any spike history prior to $t_0$.

\sssu{General form of the flow}

We start by solving the associated homogeneous Cauchy problem
\begin{equation}\label{cauchy}
\left\{
\begin{array}{cclc}
\frac{dV(t,\omega)}{dt} & = & \dtt V(t,\omega), \\
V(t_0) & = & v,
\end{array}
\right.
\end{equation}

The following theorem is standard and can be found e.g. in \cite{brockett:70}.

\begin{theorem}
If $\dtt$ is a square matrix whose elements are bounded, the sequence of matrices $M_k(t_0,t,\omega)$ defined recursively by:
$$M_0(t_0,t,\omega)=\cI_N$$
$$M_k(t_0,t,\omega)=\cI_N+ \displaystyle\int^{t}_{t_0} \dts M_{k-1}(s, t )ds, \quad t \leq t_1,$$
converges uniformly in $\left[ t_0,t_1\right] $. We note:
\beq\label{Gamma_as_limit}
\Gamma(t_0,t,\omega) \deq \lim_{k \rightarrow \infty} M_k(t_0,t,\omega)
\eeq
the limit function called ``flow". 
\end{theorem}

\sssu{Exponential flow}\label{SExpFlow}

If $\dtt$ and $\dts$ commute, $\forall s,t$, then the flow takes the form of an exponential:
%
$$
\Gamma(t_0,t,\omega)=\sum_{k=0}^{\infty} \frac{1}{k!}(\int_{t_0}^{t} \dts ds)^k=e^{\int_{t_0}^{t} \dts ds}.
$$


From (\ref{Phi}) the commutation condition reads %
$ C^{-1}\bra{\Gap - G(t,\omega)}  C^{-1}\bra{ \Gap - G(s,\omega)} = C^{-1}\bra{ \Gap - G(s,\omega)} C^{-1} \bra{ \Gap - G(t,\omega)}$. 
Since we have assumed that all neurons have the same capacity, the commutation condition reduces to  $\bra{\Gap - G(t,\omega)}  \bra{ \Gap - G(s,\omega)} = \bra{ \Gap - G(s,\omega)} \bra{ \Gap - G(t,\omega)}$, i.e. $\Gap$ and $G(t,\omega)$ commute for all $t$s. Since $G(t,\omega)$ is diagonal, the commutation condition reads $g_i(t,\omega) \Gap_{ij} \, = \, \Gap_{ij}g_j(t,\omega), \, \forall i,j$. Therefore, here are the \textit{only cases} where the commutation property holds:

\begin{enumerate}[(i)]
\item $\Gap=0$;
\item $G(t,\omega)=\kappa(t,\omega) \cI_N$ where $\kappa(t,\omega))$ is a real function.
\end{enumerate}

These cases  correspond respectively to the following examples.

\subsubsection*{No electric synapse}

In this case $\dtt=-\frac{1}{c} \, G(t,\omega)$ is a diagonal matrix. Thus, the flow takes the exponential form:
\begin{equation}\label{GammaSpe1}
\Gamma(t_0,t,\omega)= e^{\int^{t}_{t_0} \dts ds}=e^{-\frac{1}{c}\int_{t_0}^t G(s,\omega)ds} ,
\end{equation}
which is also a diagonal matrix.

\subsubsection*{Simple chemical conductances}

In this case $\dtt=\frac{1}{c} \Gap - \frac{\kappa(t,\omega)}{c} \cI_N$ and the flow takes an exponential form:
\begin{equation}\label{GammaSpe2}
\Gamma(t_0,t,\omega)=e^{-\frac{1}{c} \, \int _{t_0}^t \kappa(s,\omega) \, ds} e^{\frac{1}{c} \Gap(t-t_0)}.
\end{equation}
It is not diagonal in the canonical basis, but it can be diagonalized by an orthogonal variable change. 
%

\sssu{General form for the flow}

In the general case, namely \textit{in any model taking into account simultaneously chemical and electric synapses with a generic form},  $\Gap$ and $G(t,\omega)$ do not commute, and the flow (\ref{Gamma_as_limit})  does not read as an exponential but as a general Dyson series:
$$
\Gamma(t_0,t,\omega) = \mathcal{I}_N+\sum_{n=1}^{+\infty} \int_{t_0}^{t} \int_{t_0}^{s_1} \dots \int_{t_0}^{s_{n-1}} \Delta(s_1,\omega) \dots \Delta(s_n,\omega) \, ds_n \dots ds_1.
$$
Setting $B=C^{-1} \Gap$ and $A(t,\omega)=-C^{-1}G(t,\omega)$, this equation reads:
$$
\Gamma(t_0,t,\omega)= \mathcal{I}_N+\sum_{n=1}^{+\infty} \int_{t_0}^{t} \cdots \int_{t_0}^{s_{n-1}} \prod_{k=1}^n (B+A(s_k,\omega)) d_{s_1}\cdots d_{s_n,},
$$
where $\prod$ denotes the matrix product, hence ordered. Finally, the flow can be written:
\beq\label{GammaSeries}
\Gamma(t_0,t,\omega)= \mathcal{I}_N+ \sum_{n=1}^{+\infty} 
\sum_{
\tiny{
\begin{array}{ccc}
X_1 = \pare{B,A(s_1,\omega)}\\
X_2 = \pare{B,A(s_2,\omega)}\\
\dots\\
X_n = \pare{B,A(s_n,\omega)}\\
\end{array}
}
}
\int_{t_0}^{t} \cdots  \int_{t_0}^{s_{n-1}} \prod_{k=1}^n X_k \, d_{s_1}\cdots d_{s_n}.
\eeq
This form is quite cumbersome when $A(t,\omega),B$ do not commute. It has nevertheless the following property.

\sssu{Exponentially bounded flow}\label{S_exist_uniq}

\textbf{Definition:} An exponentially bounded flow is a two
parameter $(t_0,t)$ family $\{\Gamma(t_0,t,\omega)\}_{t \leq t_0}$ of flows such that, $\forall \omega \in \Omega$:
\begin{enumerate}
  \item $\Gamma(t_0,t_0,\omega)= \cI_N$ and $\Gamma(t_0,t,\omega)\Gamma(t,s,\omega) = \Gamma(t_0,s,\omega)$ whenever $t_0 \leq t \leq s$;
  \item For each $v \in \mathds{R}^N$ and $\omega \in \Omega$, $(t_0, t) \rightarrow \Gamma(t_0,t,\omega)v$ is continuous for $t_0 \leq t$;
  \item There is $M > 0$ and $m > 0$ such that  : 

\beq\label{Expdecay}  
  ||\Gamma(s,t,\omega)|| \leq Me^{-m(t-s)}, s \leq t.
\eeq
\end{enumerate}

Recall that a strong solution of the SDE (\ref{DNN}) is a stochastic process for which the paths are right-continuous with left limits everywhere with probability one, adapted to the filtration generated by $W(t)$. From \cite{wooster:11} we have the following theorem:

\begin{theorem}
If (\ref{Gamma_as_limit}) converges to an exponentially bounded flow $\Gamma(t_0,t,\omega)$, there is a unique strong solution for $t \geq t_0$ given by:
\begin{equation}\label{FlowDNN}
V(t_0,t,\omega) = \Gamma(t_0,t,\omega)v + \displaystyle \int_{t_0}^t \Gamma(s,t,\omega)f (s,\omega)ds + \frac{\sigma_B}{c} \displaystyle \int_{t_0}^t \Gamma(s, t,\omega)dW(s).
\end{equation}
\end{theorem} 

Thus, given an initial condition $v$ at a time $t_0$ and a noise trajectory, this equation gives the membrane potential vector at time $t$ by integration of the flow, provided $\max_{k=1 \dots N} \max_{u  \in [t_0,t]} V_k(u) < \theta$
(sub-threshold dynamics).
This is a classical form although $\Gamma$ has a complex structure (\ref{GammaSeries}) and a non trivial dependence in the raster history.\\

Let us now show that (\ref{GammaSeries}) converges to an exponentially bounded flow. 
If $\bar{G}=0$ then $\Gamma(s,t,\omega)$ given by (\ref{GammaSpe1}), is diagonal and exponentially bounded.
 In this case  $\Gamma(s,t,\omega)=diag(e^{-\frac{1}{c}\int_s^t g_k (u,\omega)du})$ where $g_k$ is given by (\ref{gktomega}). Consequently, setting $g_L=\inf_{\omega,u,k} g_k(u,\omega)$, the smallest conductance value attained when no neuron fires ever, in the absence of gap junctions, we have:
\begin{equation}\label{expboundwgj}
||\Gamma(s,t,\omega)|| \leq e^{-\frac{g_L}{c}(t-s)}.
\end{equation}
This is therefore an exponentially bounded flow. \\

 When $\bar{G} \neq 0$ we use the following perturbational result  (for details see corollary 2.2.3 in \cite{gil:05}).
Set $\hat{\Phi}(t,\omega)=-C^{-1}G(t,\omega)$. Considering $B=C^{-1}(\bar{G})$,  as a (not necessarily small) perturbation of $\hat{\Phi}(t,\omega)$, we have:


\begin{theorem}\label{Thexpst}
Let the flow $\hat{\Gamma}(s,t,\omega)$ be the exponentially bounded flow of equation (\ref{expboundwgj}), obtained when $\hat{\Phi}(t,\omega)=-C^{-1}G(t,\omega)$. Then the flow $\Gamma(s,t,\omega)$ of equation (\ref{cauchy}), when $\bar{G}\neq 0$ and $\dtt=\hat{\Phi}(t,\omega)+B$ obeys the inequality
\begin{equation}\label{expfremarbound}
\|\Gamma(s,t,\omega)|| \leq e^{-\frac{g_L}{c} (t-s)}e^{\int_s^t ||\hat{\Phi}(r,\omega)-\Phi(r,\omega)\|dr }. \nonumber
\end{equation}
Moreover,
\begin{equation}\label{expfremarbounddif}
\|\Gamma(s,t,\omega)-\hat{\Gamma}(s,t,\omega)|| \leq e^{-\frac{g_L}{c}(t-s)}(e^{\int_s^t ||\hat{\Phi}(r,\omega)-\Phi(r,\omega)\|dr }-1). 
\end{equation}
\end{theorem}

\bigskip

Using  (\ref{expboundwgj}) the flow associated to equation (\ref{DNN}) satisfies the following inequality with $||B||=b$:
\begin{equation}\label{expfremarbound2}
||\Gamma(s,t,\omega)|| \leq  e^{-\frac{g_L}{c}(t-s)}e^{b(t-s)}=  e^{(b-\frac{g_L}{c})(t-s)} 
\end{equation} 

Therefore we can ensure exponentially bounded flow for equation (\ref{DNN}) when $b<\frac{g_L}{c}$.
Since $\bar{G}$ is symmetric and $C=c \, \cI_N$ the norm $b=\|C^{-1}\bar{G}\|$ is equal to $\frac{\sigma_1}{c}$, where $\sigma_1$ is the largest eigenvalue of $\bar{G}$ and has the physical dimension of a conductance.
So, theorem \ref{Thexpst}
provides a (sufficient) condition for the existence of a strong solution, given by:

\beq\label{Condexpflow}
\sigma_1 < g_L.
\eeq

The largest eigenvalue of $\bar{G}$ has to be smaller than the leak conductance. 

\sssu{Remarks}

\begin{itemize}

\item As stated e.g. in \cite{galarreta-hestin:01} the typical electrical conductance values are of order $1$ nano-Siemens, while according e.g. to \cite{wohrer-kornprobst:09} the leak conductance of retinal ganglion cells is of order $50$ micro-Siemens. Therefore,
the condition (\ref{Condexpflow}) is compatible with the biophysical values of conductances in the retina. 

\item Looking at the series (\ref{GammaSeries}) one may think that the exponentially bounded flow is ensured whenever $\Phi(t)$ have a negative spectrum. This property \textit{is in general not} determined by the eigenvalues of $\Phi(t)$ in the non-autonomous case. There are examples in which the matrix $\Phi(t)$ have negative real eigenvalues $\forall t$, but the solutions of the corresponding differential equation grow in time. For a review and intuitive explanation see \cite{josic-rosenbaum:08} and example 3.5 in \cite{chicone-latushkin:99}. For a more complete mathematical analysis see \cite{wooster:11}. Therefore, in order to ensure a unique strong solution to (\ref{DNN}) one needs to find conditions ensuring exponentially boundedness.

\item 
The Dyson series (\ref{GammaSeries}) is obviously intractable on numerical and on analytical grounds.
A natural idea is to truncate this series. If gap junctions conductances are small compared to
chemical synapses, one can reorder the series 
writing first terms containing no term $\Gap$,
then one term $\Gap$, two terms, and so on.
The order of truncation is fixed by the norm of $\Gap$,
which is equal to its spectral radius since this is a symmetric matrix. If this radius is small enough truncation at small order in $\Gap$ provide a good approximation of (\ref{GammaSeries}). These aspects will be discussed in a separate paper.
\end{itemize}

\subsection{Flow and firing regime}\label{Ssffr}

The flow (\ref{GammaSeries}) characterizes the evolution of the membrane potential vector in the sub-threshold regime. Let us now extend its definition to the firing regime.

\sssu{Extended flow}

For a raster $\omega$, recall that $\tf{j}{r} (\omega)$ is the $r$-th time of firing of neuron $j$ in the raster $\omega$. Let $\Tf{p}$ be the  $p$-th time in which a neuron is reset, i.e. $\Tf{1}=\min_{j}\{\tf{j}{1} (\omega)\}$, and $\Tf{p}=\min_{j,r}\{\tf{j}{r} (\omega)>\Tf{p-1}\}$. Let $\K{p}{\omega}$ be the set of neurons firing at time $\Tf{p}$ i.e. $\forall j \in \K{p}{\omega}$, $\omega_j(\Tf{p})=1$.  

Consider a membrane potential trajectory \textit{compatible} with $\omega$. Thus, for an initial time $t_0<\Tf{1}$,  $v_k<\theta$, $k=1, \dots, N$. The membrane potential vector $V$ follows the evolution (\ref{FlowDNN}) until time $\Tf{1}$, where the membrane potential of neurons in $\K{1}{\omega}$ is set to $0$.
It keeps this value during the refractory period, until 
time $\Tf{1}+ \tau_{sep}$. During this period the neurons $\notin \K{1}{\omega}$ keep on evolving according to (\ref{FlowDNN}), but flow $\Gamma(t_0,s,\omega)$ undergoes a variation at time $\Tf{1}$, where conductances are updated. The smoothness of this variation depends on the assumed regularity of the $\alpha$ function (\ref{alpha}).
After the refractory period, the membrane potential of all neurons follows the evolution (\ref{FlowDNN}), until another group $\K{2}{\omega}$ fires at time $\Tf{2}$, and so on.

To take into account the refractory periods where neurons are in the rest state, we introduce a diagonal matrix $\chi \equiv \chi(s,\omega)$ with entries:
\beq\label{chi_k}
\chi_{kk}(s,\omega) \,=\,
\left\{
\begin{array}{ccc}
&0,& \quad \mbox{if} \, s \in \bigcup_{r}[\Tf{r}, \Tf{r}+\tau_{sep}], \mbox{and} \, k \in \K{r}{\omega};\\
\\
&1,& \quad \mbox{otherwise}.
\end{array}
\right.   \nonumber
\eeq 
Then we replace $\Gamma$ in (\ref{FlowDNN}) by $\Gamma\chi$. This gives
$$
V(t_0,t,\omega) = \Gamma(t_0,t,\omega) \, \chi(t,\omega) \, v + \displaystyle \int_{t_0}^t \Gamma(s,t,\omega)\, \chi(s,\omega) \, f (s,\omega)ds$$
$$
+ \frac{\sigma_B}{c} \displaystyle \int_{t_0}^t \Gamma(s, t,\omega) \, \chi(s,\omega) \, dW(s).
$$

To simplify the computations made in the section \ref{Ssts}, we add the following modelling choice. Denote $\tau_k(t,\omega) - \tau_{sep}$ the last time\footnote{By construction of the model this is an integer $\leq \ent{t}-1$ only determined by $t, \omega$.} before $t$ where neuron $k$ has been reset in the past. When $t_0 < \tau_k(t,\omega)$ we replace the 
integral $\int_{t_0}^t \Gamma(s, t,\omega) \, \chi(s,\omega) \, dW(s)$ by $\int_{\tau_k(t,\omega)}^t \Gamma(s, t,\omega) \, \chi(s,\omega) \, dW(s)$. In this way, the stochastic dependence upon the past is reset when the neuron's membrane potential is reset. On one hand, this modeling choice does not deeply change the phenomenology. On the other hand it allows us to formulate the determination of spike train statistics as a first passage problem (section \ref{STrans}). 
On the opposite, we keep the (deterministic) integral $\displaystyle \int_{t_0}^t \Gamma(s,t,\omega)\, \chi(s,\omega) \, f (s,\omega)ds$ unchanged. This term contains indeed a deep effect intrinsic to gap junctions, explained in details in section \ref{SRole}.\\

To summarize, we have thus:
\begin{eqnarray}\label{FlowDNNReset}
V(t_0,t,\omega) &=& \Gamma(t_0,t,\omega) \, \chi(t,\omega) \, v + \displaystyle \int_{t_0}^t \Gamma(s,t,\omega)\, \chi(s,\omega) \, f (s,\omega)ds \\
&&+ \frac{\sigma_B}{c} \displaystyle \int_{\tau_k(t,\omega)}^t \Gamma(s, t,\omega) \, dW(s). \nonumber
\end{eqnarray}

\sssu{Role of gap junctions on history dependence} \label{SRole}

In the case $\Gap=0$, $\Gamma$ is diagonal (eq. (\ref{GammaSpe1})) as well as $\Gamma \, \chi$. The reset of the membrane potential has the effect of removing
the dependence of $V_k(t,\omega)$ on its past since $V_k ([t] + \tau_{sep},\omega)$ is replaced by $0$. As a consequence, the $k$ component of eq. (\ref{FlowDNNReset}) holds, from the time  $\tau_k(t,\omega)$, introduced in the previous section, up to time $t$.
Therefore, eq. (\ref{FlowDNNReset}) factorizes 
as a set of $N$ equations \cite{cessac:11b}:
\begin{eqnarray}\label{FlowDNNResetcomp_k}
V_k(t,\omega) &=& \Gamma_{kk}(\tau_k(t,\omega),t,\omega) \,  v_k + \displaystyle \int_{\tau_k(t,\omega)}^t \Gamma_{kk}(s,t,\omega)\, f_k(s,\omega)ds    \nonumber \\
&& + \frac{\sigma_B}{c} \displaystyle \int_{\tau_k(t,\omega)}^t  \Gamma_{kk}(s, t,\omega) \,  dW_k(s), \end{eqnarray}
with $\Gamma_{kk}(\tau_k(t,\omega),t,\omega)=e^{-\frac{1}{c}\int_{t_0}^t g_k(s,\omega)ds}$ in agreement with (\ref{GammaSpe1}). \\

In this case, the reset has the effect of erasing the dependence of $V_k$ on its past anterior to its last firing time. Note that, in equation (\ref  {FlowDNNResetcomp_k}), the flow is integrated from the time $\tau_k(t,\omega)$, but the total conductance defining the flow (the term $\int_{t_0}^t g_k(s,\omega)ds$) corresponds to an integral starting from the initial time $t_0$. This is because, contrarily to membrane potentials, conductances \textit{are not reset} when a neuron fire.

The situation is quite more subtle when electric synapses are present, $\Gap\neq 0$. Neuron $k$ membrane potential is still reset at firing. From this time on, its evolution depends on the whole vector $V(t)$, in particular $V_j(t)$. But $V_j(t)$ depends on $V_k(s)_{s \leq t}$ via the gap junction connection. Due to this interaction type, the evolution of $V_k$ depends on its past before firing, via the membrane potential of the other neurons. 

%

\sssu{Initial conditions}

Equation (\ref{FlowDNNReset}) still depends on the initial condition $V(t_0)=v$. However, we are free to choose any $t_0 < \Tf{1}$. Especially, we can take $t_0 \to -\infty$.
This corresponds to a situation where the neural network has started to exist in a distant past (longer than all characteristic relaxation times in the system) and we observe it after transients. This corresponds to an ``asymptotic" regime which not necessarily stationary, if the external current $i^{(ext)}$ depends on time.

From the exponentially bounded flow property (\ref{Expdecay}) $\Gamma(t_0,t,\omega)v  \rightarrow 0$ as $t_0 \to -\infty$. Therefore upon taking $t_0 \rightarrow -\infty$ we may write (\ref{FlowDNNReset}) as:
%
$$
V(t,\omega) =  \displaystyle \int_{-\infty}^t \Gamma(s,t,\omega)\chi(s,\omega)f (s,\omega)ds + \frac{\sigma_B}{c} \displaystyle \int_{\tau_k(t,\omega)}^t \Gamma(s, t,\omega) \, dW(s).
$$

The integrals are well defined thanks to the exponentially bounded flow property. From now on we work with the limit $t_0 \to -\infty$. To alleviate notation we remove the variable $t_0$ in the expression of the membrane potential.

\sssu{Membrane potential decomposition}

The stochastic process $V(t,\omega)$ is therefore the sum: 
\begin{equation}\label{Vdec}
V(t,\omega)= V^{(d)}(t,\omega)+V^{(noise)}(t,\omega), \end{equation}
 of a deterministic part:
\begin{equation}\label{vdet}
V^{(d)}(t,\omega)=  \displaystyle \int_{-\infty}^t \Gamma(s,t,\omega)\, \chi(s,\omega) \, f (s,\omega)ds
= V^{(cs)}(t,\omega) \, + \, V^{(ext)}(t,\omega),  
\end{equation}
with:
\beq\label{Vcs}
V^{(cs)}(t,\omega) = \frac{1}{c} \, \int_{-\infty}^t \Gamma(s,t,\omega)\, \chi(s,\omega) \, I^{(cs)}(s,\omega) ds,
\eeq
the chemical synapses contribution to the membrane potential;
\beq\label{Vext}
V^{(ext)}(t,\omega) = \frac{1}{c} \, \int_{-\infty}^t \Gamma(s,t,\omega)\, \chi(s,\omega) \, I^{(ext)}(s,\omega) ds, 
\eeq
the external current + leak term contribution,
and a stochastic part:
\begin{equation}\label{vnoise}
V^{(noise)}(t,\omega)= \frac{\sigma_B}{c} \displaystyle \int_{\tau_k(t,\omega)}^t \Gamma(s, t,\omega) \, \, dW(s). 
\end{equation}

Thanks to the limit $t_0 \to -\infty$ which has removed the dependence in the initial condition $V^{(d)}(t,\omega)$ is uniquely determined by the spike history $\omega$ (and the time dependence of the external current $i^{(ext)}$, if any). Likewise, $V^{(noise)}(t,\omega)$ is the integral of the Wiener process with a weight $\Gamma(s, t,\omega)$ depending on the spike history.

\section{Spike train statistics and Gibbs distribution}\label{Ssts}

This section is devoted to the characterization of spike train statistics in the model. The main result establishes that spike train are distributed according to a Gibbs distribution. Note that we do not make any assumption on the \textit{stationarity} of dynamics: the present formalism affords to consider as well a time dependent external current (stimulus).   There is no explicit form of the potential determining the Gibbs distribution in the general case, but upon an assumption discussed in section \ref{SApp}, the potential can be approached by an analytic form. This form, is further discussed in the section \ref{SCons} with its deep connections, on one hand, with the Generalized Linear Model (GLM) and on the other with maximum entropy models.

For the non specialized readers, we give here the main ideas behind mathematics. Spike statistics is characterized by a family of transition probabilities giving the probability to have a spike pattern $\omega(n)$ given a past spike history $\sif{n-1}$. From this set of transitions probabilities, one defines the Gibbs distribution depending on the bio-physical parameters defining the model. This is done in a similar way as homogeneous and positive Markov chains transition probabilities define the invariant distribution of the chain, although the case considered here is more general (we do not assume stationarity and we do not assume a finite memory).

\subsection{Transition probabilities}\label{STrans}

We want to determine the probability $\Probc{\omega(n)}{\bloc{-\infty}{n-1}}$  to have a spiking pattern $\omega(n)$ at time $n$ given the spike history $\bloc{-\infty}{n-1}$. This can be stated as a first passage problem \cite{burkitt:06a,burkitt:06b,touboul-faugeras:07}. \\

Fix $\omega$, $n$ and $t < n$. Set:
\beq\label{theta_hat}
\widehat{\theta}_k(t,\omega)=\theta - V_k^{(d)}(t,\omega),
\eeq
the distance of the deterministic part of the membrane potential to the threshold.
Neuron $k$ will emit a spike at time $n$ if there exist a time $t \in [n-1,n]$ such that $V^{(noise)}_k(t,\omega)=\widehat{\theta}_k(t,\omega)$. 

Denote:
\begin{equation}\label{sigma_k}
\sigma^2_k(t,\omega)=\frac{\sigma_B^2}{c^2} \sum_{j=1}^{N} \int_{\tau_k(t,\omega)}^t \Gamma_{kj}^2(s,t,\omega) \, ds.  \nonumber  
\end{equation}
Following \cite{touboul-faugeras:07} Dubins-Schwarz' theorem can be used to change the time scale to write $V^{(noise)}_k(t,\omega)$ as a Brownian motion $V^{(noise)}_k(t,\omega) \deq W_{\sigma^2_k(t,\omega)}$ and the spiking condition at time $n$ reads:
$$W_{\sigma^2_k(t)}= \widehat{\theta}_k(t,\omega).$$
This equation characterizes the first hitting time $h_k$ of the Brownian motion $W_{\sigma^2_k(t)}$ to the``boundary" $\widehat{\theta}_k(t,\omega)$. \\

Denote $\Prob{h_1, \dots, h_N}$ the joint law of the first hitting times of neurons $1, \dots, N$. For a spiking pattern $\omega(n)$ divide the set of indices $\Set{1, \dots, N}$ into two subsets $\cS^+(n,\omega)=\Set{ k \in \Set{1, \dots,N}, \omega_k(n)=1}$ and $\cS^-(n,\omega)=\Set{ k \in \Set{1, \dots,N}, \omega_k(n)=0}$.
We have:
\beq\label{PTrans}
\Probc{\omega(n)}{\bloc{-\infty}{n-1}}
=
\Prob{\bigcap_{k \in \cS^+(n,\omega)} \Set{h_k \in [n-1,n[}
\, 
\cap \bigcap_{l \in \cS^-(n,\omega)} \Set{h_l > n}}.
\eeq

This equation can be written in terms of an integral of the join density of hitting times. 

The first passage problem can be solved in simple one dimensional situations following a method developed by Lachal \cite{lachal:98}. The Laplace transform of the first hitting time density can be obtained as
a solution of a PDE. However, we haven't been able to find a general form for the joint density of the $N$-dimensional problem in our model.

An alternative approach is to use the Fokker-Planck approach for IF models developed by several authors \cite{brunel-hakim:99,lindner-garcia:04,burkitt:06a,burkitt:06b,ostojic-etal:09}. The resulting equations are seemingly not solvable  unless using a mean-field approximation in the thermodynamic limit as done e.g. by Ostojic-Brunel-Hakim \cite{ostojic-etal:09} in the specific case where chemical and electric synapses are constant and equals. The Fokker-Planck method is a dual approach to the first passage problem  but it meets the same 
difficulties.\\

In the next section and in the appendix, we establish general conditions ensuring nevertheless that this system of transition probabilities define a Gibbs distribution. Then, in section \ref{SApp} we propose an approximation allowing an explicit computation of transition probabilities. 

\ssu{The Gibbs distribution}\label{SGibbs}

The transition probabilities (\ref{PTrans}) define
a stochastic process on the set of rasters, where the probability of having a spiking pattern $\omega(n)$ at time $n$ depends on an infinite past $\bloc{-\infty}{n-1}$. This is the extension of the concept of Markov chain to processes having an infinite memory. Such processes are called ``chains with complete connections"
\cite{onicescu-mihoc:35,maillard:07,fernandez-maillard:05}. Under suitable conditions, exposed in the appendix, a sequence of transition probabilities defines a unique probability distribution $\mu$ on the set of rasters, called a ``Gibbs distribution\footnote{Here, Gibbs distributions are considered in a general setting. In the language of statistical physics we consider infinite-range potentials and non translation-invariant potentials \cite{ruelle:69,georgii:88}. However, since we are considering time evolving processes we shall use the (equivalent) language of stochastic processes and chains with complete connections  \cite{fernandez-maillard:05,maillard:07}.}". We give here its main properties.

\sssu{Gibbs potential}\label{SPot}

The function:
\beq\label{GibbsPot}
\phi\pare{n,\omega}=\log \Probc{\omega(n)}{\bloc{-\infty}{n-1}},
\eeq
is called a \textit{Gibbs potential}. 
Then, the probability of any block $\bloc{m}{n}$, $n>m$, given the past $\sif{m-1}$ is given by:
\beq\label{pCondphi}
\Probc{\bloc{m}{n}}{\sif{m-1}} \,=\, 
e^{\sum_{l=m}^n \phi\pare{l,\omega}}.
\eeq
This form
reminds Gibbs distributions on spin lattices, with
lattice translation-invariant probability distributions given specific boundary conditions.
Given a spin-potential of spatial-range $n$, the probability of a spin block depends upon the state of the spin block, as well as spins states
in a neighborhood of that block. The conditional probability of this block
given a fixed neighborhood is the exponential of the energy characterizing physical interactions, within the block, as well as interactions with the boundaries. 
In (\ref{pCondphi}), spins are replaced by spiking patterns;
space is replaced by time.
Spatial boundary conditions are here replaced
by the dependence upon the past $\sif{m-1}$. 
The potential (\ref{GibbsPot}) has important properties:

\bit
\item \textit{It corresponds to a process with memory}, as opposite e.g. to other statistical physics approaches of spike trains statistics based on maximum entropy models of Ising type \cite{schneidman-berry-etal:06,cocco-leibler-etal:09,tkacik-prentice-etal:10} and its extensions \cite{ganmor-segev-etal:11a,ganmor-segev-etal:11b}. In those models, spike dynamics has no memory and the probability of events occurring at different times is simply the product of probability of those events. 

\item \textit{It has infinite range, corresponding to an infinite memory.} This is due on one hand to the synaptic responses (\ref{alpha}) and on the other hand to the presence of gap junctions (see section \ref{SRole}). This can be readily seen on the form of the flow (\ref{GammaSeries}). Fortunately, the exponentially bounded flow property allows to approximate the potential by a finite range potential corresponding to a Markov process (section \ref{SMarkov}).
  
\item \textit{It is normalized}, since is the log of a transition probability. Hence, there is here no need to compute the ``partition function".

\item \textit{It depends on the parameters defining the network (synaptic weights, external currents, biophysical parameters) in a nonlinear way},
as opposite to, e.g. maximal entropy models of Ising type \cite{schneidman-berry-etal:06,tkacik-prentice-etal:10} and extensions \cite{ganmor-segev-etal:11a,ganmor-segev-etal:11b}, where the potential depends linearly on parameters, but where the number of parameters grows  fast with the number of neurons, and whose interpretation is controversial.

\eit

\sssu{Gibbs distribution}

In this context, a Gibbs distribution is a probability measure $\mu$ on the set of rasters such that, 
for any block $\bloc{m}{n}$:
\beq\label{DefGibbs}
\moy{\bloc{m}{n}}=
\int_{\seq{\cA}{-\infty}{m-1}} 
e^{\sum_{l=m}^n \phi\pare{l,\omega}} \mu(d\omega).
\eeq
where the notation $\seq{\cA}{-\infty}{m-1}$ has been introduced in section \ref{SSpikeTrains}. 

The Gibbs distribution has the following properties (see
\cite{vasquez-marre-etal:12,nasser-marre-etal:12} for more details).

\bit

\item When the stochastic process defined by the transition probabilities (\ref{PTrans}) is time-translation invariant (this is the case if the external current is time-independent), the Gibbs distribution satisfies a maximum entropy principle :
it maximizes the sum ``entropy + average value of the potential". This is equivalent to free energy minimization in statistical physics. This aspect is developed in section \ref{SMaxEnt}.  

\item In the case of time-translation invariant transition probabilities depending upon a finite past (homogeneous Markov chain) the Gibbs distribution is the invariant measure of the chain (it is unique provided that transition probabilities are strictly positive). This aspect is developed in section \ref{SMarkov}.  

\item Gibbs distributions are also defined in the case where the potential is not time-translation invariant and has infinite range, upon mathematical assumptions on transition probabilities stated in the appendix.
\eit

\ssu{Approximating the Gibbs potential}\label{SApp}

The previous subsections provide mathematical results, but they do not allow explicit computations. The basic step for this would be to have an explicit form for
the Gibbs potential (resp. the transition probabilities).
As mentioned in section \ref{STrans} this is not tractable in general. We propose here an approximation.

\sssu{The stochastic term $V^{(noise)}(t,\omega)$ and compatibility conditions}\label{SCompV}

$V^{(noise)}(t,\omega)$ is the integral of a Wiener process (eq. (\ref{vnoise})). As a consequence this is a Gaussian process, with independent increments, mean $0$, and  covariance matrix:
$$
\Q(t,\omega) \deq \Cov{V^{(noise)}(t,\omega), \, V^{(noise)^T}(t,\omega)},$$
where $^T$ denotes the transpose. From standard Wiener integration we have:
$$
\Q(t,\omega)=\frac{\sigma_B^2}{c^2} \Exp{\int_{\tau_k(t,\omega)}^t \Gamma(s, t,\omega) \, \chi(s,\omega) \, dW(s) (\int_{\tau_k(t,\omega)}^t \Gamma(s', t,\omega) \, \chi(s',\omega) \, dW(s'))^T}
$$
$$=\frac{\sigma_B^2}{c^2} \int_{\tau_k(t,\omega)}^t \int_{\tau_k(t,\omega)}^t \Gamma(s, t,\omega) \, \chi(s,\omega) \chi^T(s',\omega) \Gamma^T(s', t,\omega) \, \Exp{dW(s)dW(s')^T},
$$
so that:
\beq\label{covariance}
\Q(t,\omega)=\frac{\sigma_B^2}{c^2} \int_{\tau_k(t,\omega)}^t \Gamma(s, t,\omega) \, \chi(s,\omega) \, \Gamma^T(s, t,\omega) \, ds ,
\eeq
where we used $\chi(s,\omega) \chi^T(s,\omega)=\chi(s,\omega)$.\\

However, as mentioned in section \ref{SRemarksFrame} the model definition imposes compatibility conditions between the trajectories of the membrane potential 
$V(t,\omega)$ and the raster $\omega$ (see eq. (\ref{Compatibility})).
Fixing a raster $\omega$, fixes the flow $\Gamma$ as well as the deterministic part of the membrane potential $V^{(d)}(t,\omega)$ (eq. (\ref{vdet})). Then, $V^{(noise)}(.,\omega)$ has to obey the following compatibility conditions:
\beq\label{NoiseCompatibility}
\forall k=1, \dots N, \, \forall n \leq t, \quad
\left\{
\begin{array}{lllllll}
\omega_k(n)=0 &\Leftrightarrow & \max_{u \in ]n-1,n]} \bra{V_k^{(noise)}(u,\omega) - \widehat{\theta}_k(u,\omega)}< 0;\\
\omega_k(n)=1 \, &\Leftrightarrow & \max_{u \in ]n-1,n]} \bra{V_k^{(noise)}(u,\omega) - \widehat{\theta}_k(u,\omega)} \geq 0;
\end{array}
\right.
\eeq

As a consequence, when computing the law of $V_k^{(noise)}(.,\omega)$ we have to take these constraints into account\footnote{We warmly acknowledge E. Shea-Brown for this deep remark.}. We write $\ProbC{V_k^{(noise)}(.,\omega)}$ the law of $V_k^{(noise)}(.,\omega)$ given these constraints. Although, $V^{(noise)}_k(t,\omega)$ is  Gaussian, $\ProbC{V_k^{(noise)}(.,\omega)}$ is not a Gaussian distribution. For example, if $\omega_k(n)=0$
then, necessarily, $V^{(noise)}_k(t,\omega) < \theta - V^{(d)}_k(t,\omega)$, $\ent{t}+1=n$, while a Gaussian random variable takes unbounded values.

 \sssu{A Gaussian approximation.}\label{SGauss}
 
A plausible approximation consists then of approximating 
$\ProbC{V^{(noise)}(.,\omega)}$ by the Gaussian law of $V^{(noise)}$, i.e. ``ignoring" compatibility conditions. This approximation can be justified under the following conditions.\\

\begin{itemize}

\item \textbf{Weak noise.} Consider the compatibility conditions for $\omega_k(n)=0$:
$\forall u \in ]n-1,n]$, $V_k^{(noise)}(u,\omega) < \widehat{\theta}_k(u,\omega)$. What is the probability that the Gaussian noise $V_k^{(noise)}(u,\omega)$ violates this condition for some $u$ in the interval $]n-1,n]$ ? 

Denoting $\sigma^2_k(u,\omega) \deq \Q_{kk}(u,\omega)$, the variance of $V_k^{(noise)}(u,\omega)$ at time $u$ (given explicitly by eq.  (\ref{covariance})), this probability is given by:
$$
\Prob{V_k^{(noise)}(u,\omega) \geq \widehat{\theta}_k(u,\omega)} \, = 
\, \frac{1}{\sqrt{2 \pi}} \, \int_{\frac{\widehat{\theta}_k(u,\omega)}{\sigma_k(u,\omega)}}^{+\infty}
e^{-\frac{h^2}{2}} dh.
$$

From the explicit form of $\sigma^2_k(u,\omega)$ (eq. (\ref{covariance})) and from the exponentially bounded flow property of $\Gamma$, $\sigma^2_k(u,\omega)$ is upper
bounded,  uniformly in $u$ and $\omega$, by $\frac{\sigma_B^2}{c^2} B$
where $B$ is some constant depending on the model parameters and where $\sigma_B$ is the Wiener noise intensity. Thus, $\sigma^2_k(u,\omega) \to 0$ as $\sigma_B \to 0$.

If $\sigma_B=0$ then the compatibility conditions for $\omega_k(n)=0$ is 
$\forall u \in ]n-1,n]$, $ \widehat{\theta}_k(u,\omega) >0$ (i.e. $V_k^{(d)}(u,\omega) < \theta$).  Assume that $\widehat{\theta}_k(u,\omega) > \epsilon $ for some $\epsilon >0$. 

If $\sigma_B \neq 0$, $\frac{\widehat{\theta}_k(u,\omega)}{\sigma_k(u,\omega)} > \frac{\epsilon c}{\sqrt{B}} \frac{1}{\sigma_B}$, so that:
$$
\Prob{V_k^{(noise)}(u,\omega) \geq \widehat{\theta}_k(u,\omega)} \, < 
\, \frac{1}{\sqrt{2 \pi}} \, \int_{\frac{\epsilon c}{\sqrt{B}} \frac{1}{\sigma_B}}^{+\infty}
e^{-\frac{h^2}{2}} dh= \frac{1}{2} \pare{1 - erf\pare{\frac{\epsilon c}{\sqrt{B}} \frac{1}{\sigma_B}}},
$$
where $erf(x)$
admits the following series expansion as $x \to + \infty$:
$$erf(x) = 1- e^{-x^2} \frac{1}{\sqrt{\pi}} \pare{\frac{1}{x}-\frac{1}{2x^3} + \frac{3}{4x^5} - \frac{15}{8x^7}} + \circ\pare{x^{-8}e^{-x^2}}.$$

Therefore, as $\sigma_B \to 0$, we have:
\beq
\Prob{V_k^{(noise)}(u,\omega) \geq \widehat{\theta}_k(u,\omega)} \, < 
\, \frac{1}{2} e^{-x^2} \frac{1}{\sqrt{\pi}} \pare{\frac{1}{x}-\frac{1}{2x^3} + \frac{3}{4x^5} - \frac{15}{8x^7}} + \circ\pare{x^{-8}e^{-x^2}},
\eeq
with $x=\frac{\epsilon c}{\sqrt{B}} \frac{1}{\sigma_B}$.

This shows that the probability that the noise violates the compatibility condition decreases exponentially fast as $\sigma_B \to 0$. 
As a consequence, when the noise is small, approximating 
$\ProbC{V_k^{(noise)}(.,\omega)}$ by the Gaussian law of $V_k^{(noise)}$ provides a reliable approximation. This amounts to considering that the spikes arising between $[n-1,n[$ are determined by the deterministic part of the membrane potential, not by the noise. If a neuron is about to fire at time $n$ because its deterministic part crosses the threshold at a time $t \in [n-1,n[$, the (weak) noise can affect the time $t$ where the crossing occurs, but, with high probability this time stays in the interval $[n-1,n[$.

\item \textbf{Time discretization.} Beyond the compatibility conditions, an additional aspect makes the analytic computation of transition probabilities delicate: membrane potential time evolution is continuous. Since we are focusing here on spike statistics, where spike occur on discrete times, this obstacle can be raised using the following approximation.
 We replace the spike condition of neuron $k$ at time $n$: $\exists u \in ]n-1,n], V_k^{(noise)}(u,\omega) \geq \widehat{\theta}_k(u,\omega)$ by: $V_k^{(noise)}(n,\omega) \geq \widehat{\theta}_k(n,\omega)$. The argument supporting this choice is the following. Suppose that $V_k^{(noise)}(n-1,\omega) < \widehat{\theta}_k(n-1,\omega)$; if $V_k^{(noise)}(n,\omega) \geq \widehat{\theta}_k(n,\omega)$ then with probability 1, the continuous process crosses the threshold between $]n-1,n]$, therefore the spike is registered at time $n$. On the other hand if $V_k^{(noise)}(n,\omega) < \widehat{\theta}_k(n,\omega)$ there is nevertheless a positive probability that the continuous process crosses the threshold at some time $t$ between $]n-1,n[$. In this case, there will be a spike that our approximation neglects since $V_k^{(noise)}(n,\omega) < \widehat{\theta}_k(n,\omega)$.
 The probability of occurrence of such an event can be explicitly computed for general diffusion processes (see \cite{baldi:02}) and depend on the noise intensity $\sigma_B$ and time step discretization $\delta$. It is given by: 
%
%
$$C_{\widehat{\theta}}\exp \pare{\frac{-2(\widehat{\theta}_k(n-1,\omega)- V_k^{(noise)}(n-1,\omega))(\widehat{\theta}_k(n,\omega)- V_k^{(noise)}(n,\omega))}{\delta \sigma_B}},$$
where 
$C_{\widehat{\theta}}$ is a bounded function depending on the value of $\widehat{\theta}$ at time $n-1$.  
Therefore, when $\sigma_B, \delta$ are small (as in our case) this probability is also small.

\end{itemize}

\sssu{Approached form of the potential.}\label{SAppPhi}

Thanks to these approximation we have now the following result.
If we note:
\beq\label{j_k}
\mathcal{J}_k(n,\omega) \,=\,
\left\{
\begin{array}{ccc}
&]-\infty,\widehat{\theta}_k(n-1,\omega)[,& \quad \mbox{if}\: \omega_k(n)=0;\\
\\
&[\widehat{\theta}_k(n-1,\omega),+\infty [ ,& \quad  \mbox{if}\: \omega_k(n)=1; \nonumber
\end{array}
\right.   
\eeq 
and
\begin{equation}\label{J}
\mathcal{J}(n,\omega)=\prod_{k=1}^N \mathcal{J}_k(n,\omega),
\end{equation}
where $\prod$ denotes the Cartesian product of intervals,
eq. (\ref{PTrans}) becomes, using the Gaussian approximation
for $V^{(noise)}(.,\omega)$: 
\begin{equation}\label{PTransApp}
\Probc{\omega(n)}{\bloc{-\infty}{n-1}}
=\int_{\mathcal{J}(n,\omega)} \frac{e^\frac{-V^T\Q^{-1}(n-1,\omega)V}{2}}{(2\pi)^{\frac{N}{2}}|\Q(n-1,\omega)|^{\frac{1}{2}}}dv,
\end{equation}
where $dv=\prod_{i=1}^N dV_i$.

Taking the $\log$ of (\ref{PTransApp}) we obtain the approached form of the Gibbs potential.

\sssu{Remarks}

The apparently simple form (\ref{PTransApp}) hides a real complexity. 

\begin{itemize}

\item The integration domain $\mathcal{J}(n,\omega)$ corresponds to a product of intervals involving the variables $\widehat{\theta}_k(n-1,\omega)$, the distance of $V^{(d)}_k(n-1,\omega)$ to the threshold. Now, from eq. (\ref{DNN}), (\ref{vdet}), involves the chemical synapse current $I^{(cs)}(t,\omega)$, eq. (\ref{Ics}), and
the external current $I^{(ext)}(t)$ (\ref{Iext}), integrating, via the flow $\Gamma$, the spike history 
of the network. As a consequence, the transitions probabilities depend on all parameters defining the system: synaptic weights $W_{kj}$ (eq. (\ref{Wkj}), gap junctions, external current, and biophysical parameters such as membrane capacity, or characteristic time scale of the post-synaptic potential $\alpha_{kj}(t)$ (eq. (\ref{alpha})).

%

\item  Without electric synapses the covariance (\ref{covariance}) takes a diagonal form since $\Gamma$ and $\chi$ are diagonal matrices. Thus, in this case, neurons are (conditionally upon the past), \textit{independent}.  As a consequence the transition probability $\Probc{\omega(n)}{\bloc{-\infty}{n-1}}$ factorizes:
\beq
\Probc{\omega(n)}{\bloc{-\infty}{n-1}}=
\prod_{k=1}^N \Probc{\omega_k(n)}{\bloc{-\infty}{n-1}}.
\eeq

\end{itemize}

\su{Consequences} \label{SCons}

In this section we adopt the following point of view. Assuming that the model presented here captures enough of the biophysics of real neurons, what are the -relevant for neuroscience- consequences of the mathematical results developed in the previous sections  ?  We essentially focus on spike trains analysis and argue that:

\begin{enumerate}

\item Spikes correlations are not only due to shared stimulus:
there are correlations induced by dynamics, that persists without stimulus. Moreover, \textit{in the absence of electric synapses} neurons are conditionally independent upon the past.

\item The Gibbs potential (\ref{GibbsPot}) or even its Gaussian approximation (\ref{PTransApp}), includes existing models 
for spike trains statistics such as maximum entropy models \cite{schneidman-berry-etal:06,tkacik-schneidman-etal:09,shlens-field-etal:06,shlens-field-etal:09,ohiorhenuan-mechler-etal:10,ganmor-segev-etal:11a,ganmor-segev-etal:11b,vasquez-marre-etal:12} and GLM \cite{pillow-paninski-etal:05,pillow-shlens-etal:08,pillow-ahmadian-etal:11,pillow-ahmadian-etal:11b,macke-cunningham-etal:11}.

\end{enumerate}

\ssu{Correlations structure}\label{SCorr}

\sssu{Transition probabilities do not factorize in general}

This can be illustrated in the Gaussian approximation. 
The covariance matrix $\Q(n-1,\omega)$ can always be diagonalized by an orthogonal variable change $P(n-1,\omega)$ depending on $n$ and $\omega$:
$$\Q(n-1,\omega) \, P(n-1,\omega) = P(n-1,\omega) \, \Sigma(n-1,\omega),$$
where $\Sigma(n-1,\omega)$ is diagonal with real, positive
eigenvalues $\sigma^2_k(n-1,\omega)$.

Upon the variable change $V=P(n-1,\omega) V'$, the transition probability reads therefore:
%
$$
\Probc{\omega(n)}{\bloc{-\infty}{n-1}}
=\int_{\mathcal{J'}(n,\omega)} \prod_{k=1}^N \frac{1}{\sqrt{2 \pi} \sigma_k } e^{-\frac{V'^2_k}{2 \sigma^2_k(n-1,\omega)} } \, dV_k',
$$
%
where $\mathcal{J'}(n,\omega)$ is the image of $\mathcal{J}(n,\omega)$ in the variable change. As a classical result
the variables change has transformed the join Gaussian density of $V^{(noise)}(n-1,\omega)$ into a product
of one dimensional Gaussian with mean zero and variance $\sigma^2_k(n-1,\omega)$. However, the same variable change
has transformed the domain $\mathcal{J}(n,\omega)$, reading as a product of intervals (eq. \ref{J}) into the domain $\mathcal{J'}(n,\omega)$ which is not a product any more, except in the case with no gap junctions.



Therefore, in general:
\beq
\Probc{\omega(n)}{\bloc{-\infty}{n-1}} \neq
\prod_{k=1}^N \Probc{\omega_k(n)}{\bloc{-\infty}{n-1}}.
\eeq
%


\ssu{Correlations}
	
Note first that \textit{conditional independence upon the past} does not mean \textit{independence}. Conditional independence 
upon the past means that the Gibbs potential reads:
$$
\phi_n\pare{\omega}=\sum_{k=1}^N \phi_{n,k}\pare{\omega}.
$$
This is a sum of per-neuron potentials $\phi_{n,k}$, but each of this potential is a function of the past \textit{network}
activity $\omega$.
On the opposite, neurons independence would mean that:
$$
\phi_n\pare{\omega}=\sum_{k=1}^N \phi_{n,k}\pare{\omega_k},
$$
where now the per-neuron potentials $\phi_{n,k}$ depends on 
the past activity of neuron $k$ only.

Now, what are the sources of general correlations (i.e. pairwise and higher order) in our model ? Namely, what makes the potential depend on the network history ? 
Even in the absence of electric synapses, correlations can be induced on one hand by the stimulus (the term $V^{ext}(t,\omega)$, eq (\ref{Vext})) if the external current $i^{(ext)}(t)$ in (1) has correlations between its entries ($i_k^{(ext)}(t)$  and $i_l^{(ext)}(t)$ for $k \neq l$ are correlated), and
on the other hand by the chemical synapses term $V^{(cs)}(t,\omega)$, (eq. (\ref{Vcs})). Even if the stimulus is zero, the chemical synapses term remains. We arrive then to the (somewhat obvious) conclusion that in the model the main source of correlations \textit{is not} a (shared) stimulus. It is due to interactions between neurons. \\

A deeper question is to quantify the intensity of correlations
induced by (i) chemical synapses; (ii) electric synapses; (iii)
stimulus and under which conditions (parameters value) shared stimulus correlations are dominant. This is the main additional issue to investigate neuronal encoding by a population of neurons. This requires an extended investigation beyond the scope of this paper.

\ssu{The Gibbs potential form includes existing models 
for spike trains statistics}

\sssu{Markovian approximation} \label{SMarkov}

The exponentially bounded flow property (\ref{Expdecay}) means that the norm of the flow $\Gamma(s,t,\omega)$ decay exponentially fast as $t-s$ grows. A consequence of this is the \textit{exponential decay of the variation of $\Gamma(s,t,\omega)$ with respect to $\omega$}. This property, mathematically defined in the appendix, means in words the following. Assume that we know the spike patterns of the raster $\omega$ from some integer time $m<t$ to time $t$, but we ignore the spike patterns occurring before $m$. What is the maximal error than we can make on the value of $\Gamma(s,t,\omega)$ ? As shown in the appendix, thanks to the exponential bounded flow property, this error decays exponentially fast as $t-m$ grows. The same property holds for the membrane potential. It holds as well for the Gibbs potential under conditions stated in the appendix.

As a consequence the norm of the difference between the exact Gibbs potential (\ref{GibbsPot}) and an approximate potential
where the past spike history $\bloc{-\infty}{n-1}$ is replaced by $\bloc{n-D}{n-1}$ for some integer $D>0$ called \textit{memory depth}, decreases exponentially fast with $D$. 

Therefore, we may replace the infinite range potential (\ref{GibbsPot}) corresponding to a process with infinite memory, by a truncated potential, corresponding to a Markov chain with memory depth $D$. In this case, the Gibbs distribution is the invariant probability of the corresponding Markov chain. For a small number of neurons 
this probability can be characterized using transfer matrices techniques \cite{vasquez-marre-etal:12}. For larger networks Monte Carlo methods can be used \cite{nasser-marre-etal:12}. 

\sssu{The maximum entropy principle} \label{SMaxEnt}

Assume that transition probabilities (\ref{PTrans}) are time-translation invariant, i.e. the Gibbs potential (\ref{GibbsPot}) obeys  $\phi(n+k,\omega) = \phi(n,\omega)$, $\forall k$. This corresponds to the physical concept of \textit{stationarity}. Then $\phi$ obeys a maximum entropy principle.
There is a function $\cP\bra{\phi}$, called 
\textit{topological pressure}, which is analogous of the free energy density (especially, this a generating function for cumulants) which obeys \cite{keller:98,chazottes-keller:09}:
\beq\label{MaxEnt}
\cP\bra{\phi}=\sup_{\nu \in \cM_{inv}} \pare{h\bra{\nu} \, + \, \nu\bra{\phi)}}=
h\bra{\mu} \, + \, \moy{\phi},
\eeq
where $h$ is the entropy rate (or Kolmogorov-Sinai entropy), $\cM_{inv}$ is the set of time-translation invariant probability measures, and $\mu$ is the Gibbs distribution.\\

If $\phi$ has finite range $D$: $\phi(n,\omega) \equiv \phi(\bloc{n-D}{n})$ then, a classical result from Hammersley and Clifford \cite{hammersley-clifford:71} establishes that $\phi$ can be written as:
\beq \label{expansion_phi}
\phi(\bloc{n-D}{n}) \, = \, \sum_{k=0}^{2^{ND}} \beta_k \,
\cO_k(\bloc{n-D}{n}),
\eeq
where the $\beta_k$s are real parameters, (non linear) functions of the network parameters of the model. Moreover, the
 $\cO_k$s are functions of type: 
\beq\label{Ok}
\cO_k(\bloc{n-D}{n}) = \omega_{i_1}(t_1) \dots \omega_{i_n}(t_n),
\quad i_l \in \Set{1, \dots,N}, \, t_l \in \Set{n-D, \dots, n}.
\eeq
These are characteristic functions of events: $\cO_k(\bloc{n-D}{n})=1$ whenever neuron $i_1$ fires at time $t_1$, $\dots$, neuron $i_l$ fires at time $t_l$ in the raster $\omega$. It is easy to show that there are $2^{ND}$ such functions. Note that this result is general and \textit{does not depend on the Gaussian assumption of section \ref{SGauss}}. Only the dependence of the $\beta_k$s in network parameters is constrained by the analytical form of the potential.

Using (\ref{expansion_phi}), eq. (\ref{MaxEnt}) reads:
$$
\cP\bra{\phi}=\sup_{\nu \in \cM_{inv}} \pare{h\bra{\nu} \, + \, \sum_{k=0}^{2^{ND}} \beta_k \,
\nu\bra{\cO_k(\bloc{n-D}{n})}}=
h\bra{\mu} \, + \, \sum_{k=0}^{2^{ND}} \beta_k \,
\moy{\cO_k(\bloc{n-D}{n}}.
$$\\

Assume now that we are only given a raster $\omega$ generated by our network, and that we want to approximate the exact Gibbs distribution $\mu$ from those data. A classical approach, dating back to Jaynes \cite{jaynes:57} consists of fixing a certain set of observables and to measure their empirical average value. From Hammersley-Clifford theorem any observable can be written as a linear combination of observables $\cO_k$, so there is no loss of generality in assuming that observables \textit{are} functions of the type (\ref{Ok}). Call $A_k$ the empirical average of $\cO_k$. The $A_k$s are fixed from the experimental data and therefore constitute constraints to our problem. Therefore, the maximal entropy principle (\ref{MaxEnt}) reads:
$$
\cP\bra{\phi}
=\sup_{\nu \in \cM_{inv}; \, \nu\bra{\cO_k}=A_k} h\bra{\nu} \, + \, \sum_{k=0}^{2^{ND}} \beta_k \,
A_k=
h\bra{\mu} \, + \, \sum_{k=0}^{2^{ND}} \beta_k \,
A_k.
$$
The Gibbs distribution $\mu$ \textit{maximizes the entropy rate under the constraints $\moy{\cO_k}=A_k$}.
These constraints fix the value of the $\beta_k$s in a unique way.\\

Assume now that we are not able to measure the empirical average of all $\cO_k$s: as the number of events implied in the product (\ref{Ok}) increases its probability becomes more and more difficult to estimate. So, typically, one has only a subset of constraints. The application of Jaynes' strategy consists of maximising the entropy rate under this subset of constraints. The Gibbs distribution obtained this way is only an approximation of $\mu$. 

The Jaynes strategy has been applied by many authors
to the analysis of spike train statistics, especially in the retina \cite{schneidman-berry-etal:06,tkacik-schneidman-etal:09,shlens-field-etal:06,shlens-field-etal:09,ohiorhenuan-mechler-etal:10,ganmor-segev-etal:11a,ganmor-segev-etal:11b,vasquez-marre-etal:12}. In particular, fixing firing rates and the probability of pairwise coincidences of spikes leads
to a Gibbs distribution having the same form as the Ising model.
This idea has been introduced by Schneidman et al in 
\cite{schneidman-berry-etal:06}.
They reproduce accurately the probability of
\textit{spatial} spiking pattern. Since then, their approach has known a great success
(see e.g. \cite{shlens-field-etal:06,cocco-leibler-etal:09,tang-jackson-etal:08}), although some authors raised
solid objections on this model \cite{roudi-nirenberg-etal:09,shlens-field-etal:09,tkacik-schneidman-etal:09,ohiorhenuan-mechler-etal:10} while several papers have pointed out the importance of \textit{temporal} patterns of activity at the network level \cite{lindsey-morris-etal:97,villa-tetko-etal:99,segev-baruchi-etal:04}. 
As a consequence, a few authors \cite{marre-boustani-etal:09,amari:10,roudi-tyrcha-etal:09}
have attempted to define time-dependent models of Gibbs distributions where constraints include time-dependent correlations
between pairs, triplets and so on \cite{vasquez-vieville-etal:11}. As a matter of fact, the analysis
of the data of \cite{schneidman-berry-etal:06} with such models describes more accurately the statistics
of \textit{spatio-temporal} spike patterns  \cite{vasquez-marre-etal:12}.\\

Now, the inspection of the Gibbs potential in our model  leads to several strong conclusions:

\ben

\item The Gibbs potential is definitely not Ising, and is actually quite far from an Ising model. The main reason for that is that Ising model involves only instantaneous (pairwise) events, with no memory effects. On the opposite, strong memory effects exist in our model, requiring to consider transition probabilities depending on spike history. 

\item The exact expansion (\ref{expansion_phi}) involves $2^{ND}$ constraints, making rapidly intractable any numerical methods attempting to match all constraints.
Obviously, one can argue that some $\beta_k$s can be close to $0$ so that the corresponding constraints can be ignored in the approximation of $\mu$. Unfortunately, this argument does not tell us \textit{which are} the negligible terms. It might well be that the answer depend on the network parameters as well. This aspect is under current investigations.

\item Although the expansion (\ref{expansion_phi}) of a potential into a linear combination of observables is quite natural in the realm of statistical physics and Jaynes principle, it may not be appropriate for the study of neural networks of the type studied here. Indeed, the parameters $\beta_k$s, whose number increases exponentially fast with the number of neurons, are \textit{redundant}: they are functions of the network parameters, such as synaptic weights, whose number increases, at most, like $N^2$. Using approximations  
like the Gaussian one allows to obtain an explicit form for the potential as a function of these parameters. 
One obtains in the end a nonlinear function, but with quite a bit less parameters to tune; additionally those parameters have a straightforward biophysical interpretation. This constitutes an alternative strategy to estimate spike statistics from empirical data.  
\een

\sssu{The GLM model}

The Linear-Nonlinear (LN) models and Generalized Linear Models (GLM)
 \cite{brillinger:88, mccullagh-nelder:89, paninski:04,truccolo-eden-etal:05,pillow-paninski-etal:05,pillow-shlens-etal:08,pillow-ahmadian-etal:11,pillow-ahmadian-etal:11b} constitute  prominent alternatives to MaxEnt models.  We focus here on the GLM.

Let $x \equiv x(t)$ be a time-dependent stimulus. In response to $x$ the network emits a spike train response $r$. This response does not only depend on $x$, but also on the network history $\mathcal{H}$. The GLM (and LN) assimilate the response $r$ as an inhomogeneous Poisson process: the probability that neuron $k$ emits a spike between $t$ and $t+dt$ is given by $\lambda_k(t) \, dt$, where $\lambda_k(t)$ is called ``conditional intensity".  
In the GLM  this function is given by:
\beq\label{lambdaGLM}
\lambda_k(t) = f\pare{b_k + (K_k \ast x)(t) + \sum_{j} (H_{kj} \ast r_j)(t)},
\eeq
where $f$ is a non linear function (an exponential or a sigmoid); $b_k$ is some constant fixing the baseline firing rate of neuron $k$; $K_k$ is a causal, time-translation invariant, linear convolution kernel that mimics a linear receptive field of neuron $k$; $\ast$ is the convolution product; $H_{kj}$ describes possible excitatory or inhibitory post spike effects of the $j$ th observed neuron on the $k$ th. As such, it depends on the past spikes, hence on $\omega$.
The diagonal components $H_{kk}$ describe the post spike feedback of the neuron to itself,
and can account for refractoriness, adaptation and burstiness depending on their shape. \\


 The GLM postulates that, given the history $\mathcal{H}$ and stimulus $x$, the neurons are independent (\textit{conditional independence upon the past and stimulus}). The network response at time $n$ is a spiking pattern $\omega(n)$ while the history is the spike activity $\sif{n-1}$. As a consequence of the conditional independence assumption the probability of a spiking pattern given the history of the network reads:
\beq\label{PGLM}
\Probc{\omega(n)}{\bloc{-\infty}{n-1}} = \prod_{k=1}^N \lambda_k(n)^{\omega_k(n)} (1-\lambda_k(n))^{1-\omega_k(n)},
\eeq
and the Gibbs potential is:
\beq\label{phiGLM}
\phi_n(\omega)=\sum_{k=1}^N \bra{\omega_k(n) \, \log \lambda_k(n)^{} + (1-\omega_k(n)) \log (1-\lambda_k(n))}.
\eeq
It is normalized by definition. \\

Although the form of $\lambda_k(n)$ is postulated, the GLM has been proved quite efficient to analyse retina spike trains.  In our model, one can have an explicit form for $\lambda_k(n)$ in the case without gap junctions and under the Gaussian approximation. It has the form (\cite{cessac:11a}):
\beq\label{phiGIF}
\lambda_k(\omega)=\pi(X_k(n-1,\omega)),
\eeq
where
$$\pi(x)=\frac{1}{\sqrt{2 \pi}}\int_x^{\infty}e^{\frac{-u^2}{2}}du,$$
is the sigmoid function, whereas
$
 \quad X_k(n-1,\omega)=\frac{\widehat{\theta}_k(n-1,\omega)}{\sigma_k(n-1,\omega)}.$
There is a clear similarity between equation (\ref{phiGIF}) and (\ref{phiGLM}) where $f$ in (\ref{lambdaGLM}), usually an exponential\footnote{Note that the exponential form is used for its convexity property, useful to learn the parameters of the GLM \cite{pillow-ahmadian-etal:11}.}, is replaced by $\pi$, a sigmoid. In our model the stimulus and the history are contained in the term $X_k(n-1,\omega)$ respectively via $V^{(ext)}(n-1,\omega)$ (eq. (\ref{Vext})), the external current contribution, and
$V^{(cs)}(n-1,\omega)$ (eq. \ref{Vcs})), the chemical synapses contribution to the membrane potential. Therefore, here the terms $K_k \ast x$ and $\sum_{j} H_{kj} \ast r_j$ in eq. (\ref{lambdaGLM}) are made explicit.
Note that we have nowhere specified the explicit form of the external current $i^{(ext)}$: it can have the form $(K_k \ast x)(t)$ including, e.g. the integrated effect of an input $x$, filtered by neurons layers before arriving to our model, modelling a final layer. The term $\sum_{j} H_{kj} \ast r_j$ corresponds, in our case, to the effect of chemical synapses interactions.\\

As stated several times in this paper, the conditional independence assumption does not hold any more in the presence of gap junctions. 
So, in our model with gap junctions, the GLM is an incorrect model.

\section{Conclusion}

In this work we have analyzed the join effects of 
spike history dependent chemical synapses conductances
and gap junctions in an Integrate and Fire model.
We have pointed out several technical and conceptual difficulties mainly lying in the fact that, in general,
the chemical conductance matrix $G(t,\omega)$ and
the gap junctions matrix $\Gap$ do not commute.
As we showed, this has no impact on the well posedness
of the model, provided that the values of chemical and electrical conductances are compatible with biophysics.
There exists a strong solution of the stochastic differential equations. However, except in some specific cases, the flow reads as a Dyson series, hardly tractable.
We discuss some approximations below.

We have also considered the statistics of spike trains in this model and showed that it is characterized by Gibbs distribution, time-dependent (non stationary)  whenever the external current is time-dependent. The corresponding potential can be computed under a Gaussian approximation: it has infinite range with exponentially decaying interactions.

 The main observation resulting from our analysis is that spike statistics is indecomposable. The probability of spike patterns does not
factorize as a product of marginal, per-neuron, distributions. This effect is enhanced by the presence of gap junctions. As a consequence, in that model, \textit{there is absolutely now way to defend that neurons act as independent sources}. Additionally, correlations mainly result from the chemical and electrical interactions
between neurons (correlations persist even if there is no external current / stimulus). Our work suggests especially that electric synapses could have a strong influence in spike train statistics of biological neural systems, especially the
retina where gap junctions connections are ubiquitous. \\

There are ongoing debates on the correlations between
ganglion cells in the retina: are ganglion cells independent encoders or do they act in a correlated way \cite{nirenberg-latham:98,nirenberg-latham:03,roudi-nirenberg-etal:09,mastronarde:83,schneidman-berry-etal:06,ganmor-segev-etal:11a,ganmor-segev-etal:11b}.
What are the sources of correlations \cite{panzeri-schultz:01,schneidman-bialek-etal:03} ?
Are they only due to a shared stimulus or shared noise, or
do they also result from interactions between neurons \cite{tyrcha-roudi-etal:12} ?

Our model  involves only firing cells while most cell type in the retina (amacrine, horizontal, bipolar) do not ``spike". Nevertheless, the conclusion of our paper raises questions about retina spike trains: if
dynamical correlations are so important in this simple model, how can they become weaker in a more complex neural system like the retina ? 
If such correlations exist why are they so difficult to exhibit in experiment, without controversy \cite{schneidman-berry-etal:06,roudi-nirenberg-etal:09} ? Concerning this last question, note that our spike trains are not \textit{binned}.
Binning data at a characteristic time scale of $10-20$ ms,
i.e. larger than the characteristic time scale of gap junctions and of order the time decay of chemical synapses, could have a strong impact on statistical inference. The effect of binning in our model needs however to be further investigated.

We have also pointed out that the Gibbs potential obtained in our model is quite more complex than Ising model or
similar models used in retina spike train analysis \cite{schneidman-berry-etal:06,tkacik-prentice-etal:10,ganmor-segev-etal:11a,ganmor-segev-etal:11b,tyrcha-roudi-etal:12}. Especially, it involves spatio-temporal spike patterns. Handling spatio-temporal events in Gibbs distributions requires more subtle algorithms than simultaneous events as those considered in the Ising model \cite{schneidman-berry-etal:06,cocco-leibler-etal:09,tkacik-prentice-etal:10} and its extensions \cite{ganmor-segev-etal:11a,ganmor-segev-etal:11b}. These aspects have been developed in \cite{vasquez-marre-etal:12,nasser-marre-etal:12}.

As mentioned in the introduction, one of the main motivation of this work was to better understand how ganglion cells in the retina supporting both chemical and electric synapses coordinate spatio-temporal spike patterns to encode information conveyed to the brain. Developing a further understanding
of the regulation of gap junctions, as well as the dynamic relationship between electrical and chemical transmission,
is an important challenge for the future \cite{bloomfield:09}. Although our results are still largely lying on a theoretical ground, we hope that this paper will help the neuroscience community to make a step forward in this direction.

\bigskip
\textbf{Acknowledgments}
This work was supported by the INRIA, ERC-NERVI number 227747, KEOPS ANR-CONICYT and from the European Union Seventh Framework Programme (FP7/2007- 2013) under grant agreement no. 269921 (BrainScaleS). R Cofr\'e  is funded by the French ministry of Research and University of Nice (EDSTIC). We greatly acknowledge E. Shea-Brown and E. Tanr\'e for illuminating comments.

\su{Appendix: Existence and uniqueness of a Gibbs distribution}

%

We give here the mathematical definition and proof of existence of a Gibbs distribution in our model. We also comment under which conditions the  uniqueness could be proved. 
For mathematical references on Gibbs distributions  and chains with complete connections
see e.g. \cite{georgii:88,fernandez-maillard:05,maillard:07}.

\ssu{Definitions}

For $n \in \setZ$, we note $\seq{\cA}{-\infty}{n-1}$ the set of sequences $\sif{n-1}$
and $\cF_{\leq n-1}$ the related $\sigma$-algebra, while $\cF$ is the $\sigma$-algebra related with $\Omega$. $\cP(\Omega,\cF)$ is the set of probability measures on $(\Omega,\cF)$.

\sssu*{System of transition probabilities}

A system of transition probabilities is a family $\Set{P_n}_{n \in \setZ}$ 
of functions $\Probc{ \, }{ \, } : \cA \times \seq{\cA}{-\infty}{n-1}  \to [0, 1],$
such that for every $n \in \setZ$, 

\bit 
\item for every $\omega(n) \in \cA$ ,
$\Probc{\omega(n)}{.}$ is measurable with respect to $\cF_{\leq n-1}$;
\item for every $\sif{n-1} \in \seq{\cA}{-\infty}{n-1}$, $ \sum_{\omega(n) \in \cA} \Probc{\omega(n)}{\sif{n-1}}=1$.
\eit

\sssu*{Chain with complete connections}

A chain with complete connections is 
a probability measure $\mu$ in $\cP(\Omega,\cF)$ such that
for all $n \in \setZ$ and all $\cF_{\leq n}$-measurable functions $f$:
$$\int f\left(\seq{\omega}{-\infty}{n}\right) \mu(d\omega) = \int \sum_{\omega(n) \in \cA}
f\left(\sif{n-1} \omega(n) \right) \Probc{\omega(n)}{\sif{n-1}} \mu(d\omega).
$$

\sssu*{Continuity with respect to a raster} 

We note, for $n \in \setZ$, $m \geq 0$, and $r$ integer:
$$\omega \eg{m,n} \omega' \quad \mbox{if} \quad\omega(r)=\omega'(r), \, \forall r \in \Set{n-m, \dots, n}.
$$
A  function $f(t,\omega)$ is \textit{continuous} with respect to $\omega$ if its $m$-\textit{variation}:
$$
\vm{f(t,.)}=\sup\Set{ \, | \,f(t,\omega)-f(t,\omega')\, | \,: \omega \eg{m,\ent{t}} \omega' }.
$$
tends to $0$ as $m \to +\infty$.

\sssu*{Gibbs measure} 

A probability measure $\mu$ in $\cP(\Omega,\cF)$
is a  Gibbs measure if:

\bit
\item $\mu$ is a chain with complete connection with respect to the system $\Set{\Probc{\omega(n)}{\sif{n-1}}}$.
\item $\forall n  \in \setZ$, $\forall \sif{n-1} \in \seq{\cA}{-\infty}{n-1}$, $\Probc{\omega(n)}{\sif{n-1}}>0$.
\item  The system $\Set{\Probc{\omega(n)}{\sif{n-1}}}$ is continuous with respect $\omega$.
\eit

\ssu{Existence of a Gibbs measure}

This is established by a standard result in the frame of chains with complete connections stating that a system of continuous transition probabilities on a compact space has at least one probability measure consistent with it \cite{maillard:07}. 
We give therefore here the proof of the continuity of $\Probc{\omega(n)}{\sif{n-1}}$ with respect to $\omega$. To prove this it is sufficient to show that  $\Probc{\omega(n)}{\sif{n-1}}$ depends continuously on the only functions $V^{(d)}(n-1,\omega)$ and $\mathcal{Q}(n-1,\omega)$ and that these functions are continuous with respect to $\omega$.

\sssu{$\Probc{\omega(n)}{\sif{n-1}}$ depends continuously on  $V^{(d)}(n-1,\omega)$ and $\mathcal{Q}(n-1,\omega)$}

This property is straightforward in the Gaussian approximation.
For more general diffusion processes the argument is based on the differentiability, and, consequently, on the continuity of the first hitting time density with respect to time dependent boundary perturbations \cite{costantini-etal:06}. This result holds  for multidimensional correlated processes, under conditions ensuring uniqueness of strong solutions (exponentially bounded flow in our case). 
In our case, the boundary depends continuously on $V^{(d)}(t,\omega)$ (eq. \ref{theta_hat}) thus, the continuity of the first hitting time density and therefore of the transition probabilities (eq. \ref{PTrans}) with respect to boundary perturbations implies the  continuity  of the transition probabilities $\Probc{\omega(n)}{\sif{n-1}}$ with respect to $\omega$, if $V^{(d)}(t,\omega)$ is continuous with respect to $\omega$. Additionally, in our case, the covariance $\mathcal{Q}(t,\omega)$ of the process $V^{(noise)}$ depends also on $\omega$ so we have to check that $\mathcal{Q}(t,\omega)$ is continuous  w.r.t $\omega$.

\sssu{Useful notations and results}

In the next section we prove the continuity of $V^{(d)}(t,\omega)$ and $\mathcal{Q}(t,\omega)$ w.r.t $\omega$.

For this, we introduce the following (Hardy) notation: if a function $f (t)$ is bounded from
above, as $t \rightarrow + \infty$, by a function $g(t)$ we write: $f (t) \preceq g(t)$. 

We use the following bound obtained in \cite{cessac:11b}.
If the alpha profile has the form (\ref{alpha}) where $h(t)$ is a polynomial of degree $d$, then:
\beq\label{var_alpha}
var_m [\alpha_{kj}(s,.)] \preceq  2P_d(\frac{m}{\tau_{kj}})e^{-\frac{m}{\tau_{kj}}},
\eeq
where $P_d$ is a polynomial of degree $d$, and:
\beq\label{var_g}
var_m [g_k(s,.)] \preceq  2 \sum_{l=1}^N G_{kl}P_d (\frac{m}{\tau_{kl}}) e^{-\frac{m}{\tau_{kl}}},
\eeq
uniformly in $s$.

\subsubsection{Continuity of $V_k^{(cs)}$ with respect to $\omega$}

Fix $t$, $m \geq 0$ and $\omega,\omega'$ with $\omega \eg{m,\ent{t}} \omega'$. From (\ref{Wkj}), (\ref{Vcs}), we have:
$$
\abs{V_k^{(cs)}(t,\omega)-V_k^{(cs)}(t,\omega ')} =
$$
$$ 
\abs{
\frac{1}{c}  \int_{-\infty}^t \sum_{j=1}^N 
\bra{
\Gamma_{kj}(s, t,\omega)\chi_{jj}(s,\omega)I_j^{(cs)}(s,\omega)
\, - \, \Gamma_{kj}(s, t,\omega ') \chi_{jj}(s,\omega')I_j^{(cs)}(s,\omega')
}
ds
}
$$
$$
\leq \frac{1}{c} \sum_{j,j'=1}^N |W_{jj'}| \int_{-\infty}^t  \abs{
\Gamma_{kj}(s, t,\omega)\chi_{jj}(s,\omega)\alpha_{jj'}(s,\omega) \, -\,
\Gamma_{kj}(s, t,\omega ') \chi_{jj}(s,\omega')\alpha_{jj'}(s,\omega ')} ds
$$
$$\leq 
\frac{1}{c} \sum_{j,j'=1}^N |W_{jj'}| 
\int_{-\infty}^t  
var_m [\Gamma_{kj}(s, t,.)\chi_{jj}(s,.) \alpha_{jj'}(s,.)] ds.$$

We have:
\begin{eqnarray*}
|var_m [\Gamma_{kj}(s, t,.)\chi_{jj}(s,.) \alpha_{jj'}(s,.)] &\leq&   var_m [\Gamma_{kj}(s, t,.)\chi_{jj}(s,.)] \, \sup_{\omega \in \Omega} \alpha_{jj'}(s,\omega)\\
&&  + \sup_{\omega \in \Omega} [\Gamma_{kj}(s, t,\omega)\chi_{jj}(s,\omega)] \, var_m [\alpha_{jj'}(s,.)],  
\end{eqnarray*}
and
$$
var_m [\Gamma_{kj}(s, t,.)\chi_{jj}(s,.)]\leq var_m [\Gamma_{kj}(s, t,.)]\sup_{\omega \in \Omega} \chi_{jj}(s,\omega)+ \sup_{\omega \in \Omega} \Gamma_{kj}(s, t,\omega) \, var_m [\chi_{jj}(s,\omega)]. 
$$


From equation (\ref{expfremarbounddif}) with $\hat{\Phi}(t,\omega)=B+A(t,\omega)$ and since $A(t,\omega)=\frac{1}{c}G(t,\omega)$ is diagonal we obtain the following bound:
$$|\Gamma_{kj}(s, t,\omega)-\Gamma_{kj}(s, t,\omega')| \leq \|\Gamma(s, t,\omega)-\Gamma(s, t,\omega')\| \leq e^{(b-\frac{g_{L}}{c})(t-s)} \, \bra{e^{\frac{1}{c} \, \int_s^t \|G(r,\omega)-G(r,\omega')\|dr }-1},$$
with:
%
%
$$
\|G(r,\omega)-G(r,\omega')\|  \leq N \,  \max_k var_m [g_{k}(r,.)] \deq N \, var_m [g_{\hat{k}}(r,.)].
$$
Therefore,
$$
var_m [\Gamma_{kj}(s, t,.)] \leq e^{(b-\frac{g_{L}}{c})(t-s)} \bra{e^{\frac{N}{c} \, \int_s^t  \, var_m[g_{\hat{k}}(r,.)] dr}\, - \, 1}.
$$

We have;
$$
\int_s^t  \, var_m[g_{\hat{k}}(r,.)] dr
\, \preceq \,  2 \, (t-s) \, \sum_{l=1}^N G_{\hk l}P_d (\frac{m}{\tau_{\hk l}}) e^{-\frac{m}{\tau_{\hk l}}},
$$
which converges to $0$ as $m \to +\infty$, so, that:
$$e^{\frac{N}{c} \,\int_s^t  \, var_m[g_{\hat{k}}(r,.)] dr}-1 \, \preceq \, 2 \, \frac{N}{c}  \, (t-s) \, \sum_{l=1}^N G_{\hk l}P_d (\frac{m}{\tau_{\hk l}}) e^{-\frac{m}{\tau_{\hk l}}}.$$
Therefore,
\beq\label{VarGamma}
var_m [\Gamma_{kj}(s, t,.)] \preceq 
2 \, \frac{N}{c} \, (t-s)  \,
e^{(b-\frac{g_{L}}{c})(t-s)} 
\, \sum_{l=1}^N G_{\hk l}P_d (\frac{m}{\tau_{\hk l}}) e^{-\frac{m}{\tau_{\hk l}}}.
\eeq

As shown in \cite{cessac:11b} $\alpha_{jj'}(s,\omega)$ is bounded $\forall j,j'$ by a constant $\alpha^+ < \infty$ while, following the definition of $\chi(s,\omega)$ given in section \ref{Ssffr}, $\sup_{\omega \in \Omega} \, \chi(s,\omega)=1$. Moreover, $\sup_{\omega \in \Omega} \Gamma_{kj}(s, t,\omega) \leq \sup_{\omega \in \Omega}  \| \Gamma(s, t,\omega)\|$.
Therefore:
$$
var_m [\Gamma_{kj}(s, t,.) \chi_{jj}(s,.) \alpha_{jj'}(s,.)]$$
$$
\leq \bra{var_m [\Gamma_{kj}(s, t,.)] + \sup_{\omega \in \Omega}  \, \|\Gamma(s, t,\omega)\|var_m [\chi_{jj}(s,.)]} \, \alpha^+
  + \sup_{\omega \in \Omega} \|\Gamma(s, t,\omega \, \|var_m [\alpha_{jj'}(s,.)],  
$$
and applying inequalities (\ref{expfremarbound2}) and (\ref{var_alpha}) we obtain:
$$
var_m [\Gamma_{kj}(s, t,.)\chi_{jj}(s,.)  \alpha_{jj'}(s,.)] \preceq $$
$$
2e^{(b-\frac{g_{L}}{c})(t-s)} \bra{
\alpha^+ \, \frac{N }{c} (t-s)\sum_{l=1}^N G_{\hk l} P_d(\frac{m}{\tau_{\hk l}})e^{-\frac{m}{\tau_{\hk l}}} \, +
\,  \, \frac{\alpha^+}{2}var_m [\chi_{jj}(s,.)] \, + \,
P_d(\frac{m}{\tau_{jj'}})e^{-\frac{m}{\tau_{jj'}}}  
}.
$$

Therefore:
$$
\abs{V_k^{(cs)}(t,\omega)-V_k^{(cs)}(t,\omega ')}
\preceq
\frac{1}{c} \sum_{j,j'=1}^N |W_{jj'}| 
[$$
$$
 2P_d(\frac{m}{\tau_{jj'}})e^{-\frac{m}{\tau_{jj'}}} \int_{-\infty}^t e^{(b-\frac{g_{L}}{c})(t-s)}ds \,+
 $$
 $$ 
2\alpha^+ \frac{N}{c}\sum_{l=1}^N G_{\hk l} P_d(\frac{m}{\tau_{\hk l}})e^{-\frac{m}{\tau_{\hk l}}}  \int_{-\infty}^t (t-s) e^{(b-\frac{g_{L}}{c})(t-s)}ds \,+
$$
$$
\alpha^+\int_{-\infty}^t e^{(b-\frac{g_{L}}{c})(t-s)}var_m [\chi_{jj}(s,.)]ds \, ].
$$

The integrals are convergent when $b-\frac{g_{L}}{c}<0$, i.e. when the condition (\ref{Condexpflow}) is fulfilled. In this case: 
$$ \abs{V_k^{(cs)}(t,\omega)-V_k^{(cs)}(t,\omega ')}
\preceq \frac{1}{c} \sum_{j,j'=1}^N |W_{jj'}| $$
$$
\bra{
\frac{2}{(\frac{g_{L}}{c}-b)}P_d(\frac{m}{\tau_{jj'}})e^{-\frac{m}{\tau_{jj'}}} \,+
\frac{2\alpha^+ \, N}{c(b-\frac{g_{L}}{c})^2} \sum_{l=1}^N G_{\hk l} P_d(\frac{m}{\tau_{\hk l}})e^{-\frac{m}{\tau_{\hk l}}}  \, +
\alpha^+ \int_{-\infty}^t  e^{(b-\frac{g_{L}}{c})(t-s)}var_m [\chi_{jj}(s,.)]ds }.
$$

Taking $m \rightarrow + \infty$ the two first terms converge to $0$ exponentially fast. Considering the third term we have:
$$
\int_{-\infty}^t  e^{(b-\frac{g_{L}}{c})(t-s)}var_m [\chi_{jj}(s,.)]ds$$
$$
=
\int_{-\infty}^{t-m}  e^{(b-\frac{g_{L}}{c})(t-s)}var_m [\chi_{jj}(s,.)]ds
+
\int_{t-m}^{t}  e^{(b-\frac{g_{L}}{c})(t-s)}var_m [\chi_{jj}(s,.)]ds.
$$
The second term is zero since $var_m [\chi_{jj}(s,.)]=0$ on this interval, by definition. For the first term, $var_m [\chi_{jj}(s,.)]\leq 2$ so that:
$$
\int_{-\infty}^{t-m} e^{(b-\frac{g_{L}}{c})(t-s)}var_m [\chi_{jj}(s,.)]ds
\leq
\frac{2}{\frac{g_{L}}{c}-b}  
\, e^{(b-\frac{g_{L}}{c})  \, m},
$$
which converges to $0$ exponentially fast as $m \to +\infty$
if $b-\frac{g_{L}}{c} <0$ (condition (\ref{Condexpflow})).\\

Therefore, $V^{(d)}$ is continuous with respect to $\omega$
 with an exponentially fast decay of its variation.

\subsubsection{Continuity of $V_k^{(ext)}(t,\omega)$  with respect to $\omega$}

Let us define $\tau_L=\frac{c}{g_{L}}$. 
From (\ref{Vext}) we have, for all $k$:
$$\abs{V_k^{(ext)}(t,\omega)-V_k^{(ext)}(t,\omega ')}= |\int_{-\infty}^t \sum_{j=1}^N \Gamma_{kj}(s, t,\omega)\chi_{jj}(s,\omega)\bra{\frac{E_L}{\tau_L}+\frac{1}{c}I_j^{(ext)}(s)}ds$$
$$\quad - \int_{-\infty}^t \sum_{j=1}^N \Gamma_{kj}(s, t,\omega ')\chi(s,\omega')\bra{\frac{E_L}{\tau_L} +\frac{1}{c}I_j^{(ext)}(s)}ds|$$

Without loss of generality we may consider that the external current is uniformly bounded: $\exists i^+ < +\infty$ such that,  $\forall j$, $\forall s$, $\abs{I_j^{(ext)}(s)}  \leq i^+$. Then,
$$\abs{V_k^{(ext)}(t,\omega)-V_k^{(ext)}(t,\omega ')}$$
$$\leq \bra{\frac{E_L}{\tau_L} +\frac{i^+}{c}}  \int_{-\infty}^t \sum_{j=1}^N  |\Gamma_{kj}(s, t,\omega)\chi_{jj}(s,\omega)-\Gamma_{kj}(s, t,\omega ')\chi_{jj}(s,\omega')|ds$$
$$\leq \bra{\frac{E_L}{\tau_L} +\frac{i^+}{c}}  \int_{-\infty}^t \sum_{j=1}^N  var_m [\Gamma_{kj}(s,t,.)\chi_{jj}(s,\omega)] ds$$
$$\leq \bra{\frac{E_L}{\tau_L}+\frac{i^+}{c}} \int_{-\infty}^t  \sum_{j=1}^N var_m [\Gamma_{kj}(s,t,.)] + \|\Gamma(s,t,.)\| var_m [\chi_{jj}(s,\omega)] ds,$$
from the bounds on $var_m [\Gamma_{kj}(s,t,.)\chi_{jj}(s,\omega)]$ found in the previous section. Therefore, from (\ref{VarGamma}),
$$\abs{V_k^{(ext)}(t,\omega)-V_k^{(ext)}(t,\omega ')}$$
$$\leq \bra{\frac{E_L}{\tau_L} +\frac{i^+}{c}} \bra{2 \, \frac{N}{c}\sum_{l=1}^N G_{kl} P_d(\frac{m}{\tau_{kl}})e^{-\frac{m}{\tau_{kl}}} \, \int_{-\infty}^t  (t-s) \, e^{(b-\frac{g_{L}}{c})(t-s)} \, ds   + \int_{-\infty}^t \|\Gamma(s,t,.)\| var_m [\chi_{jj}(s,\omega)] \, ds}.$$
Using the same arguments as in the previous section we get that $V_k^{(ext)}$ is continuous with respect to the raster $\omega$ with an exponentially fast convergence of its variation.

\subsubsection{Continuity of $\mathcal{Q}(t,\omega)$ with respect to $\omega$}
We have:
$$var_m \mathcal{Q}(t,.)
\leq\frac{\sigma_B^2}{c^2} \int_{\tau_k(t,\omega)}^t \| \Gamma(s, t,\omega) \, \chi(s,\omega) \, \Gamma^T(s, t,\omega)\| \, ds$$
$$\leq \frac{\sigma_B^2}{c^2} \int_{\tau_k(t,\omega)}^t 2 \| \Gamma (s,t,.)\| var_m [\Gamma(s,t,.)] \| \, ds$$
$$\leq\frac{\sigma_B^2}{c^2} \int_{\tau_k(t,\omega)}^t   2e^{2(b-\frac{g_{L}}{c})(t-s)}(e^{\frac{var_m[G(t,.)]}{c}(t-s)}-1)ds,$$
$$\leq\frac{\sigma_B^2}{c^2} \int_{-\infty}^t   2e^{2(b-\frac{g_{L}}{c})(t-s)}(e^{\frac{var_m[G(t,.)]}{c}(t-s)}-1)ds,$$

The last inequality comes from the fact that the argument of the integral is always positive. Then $var_m \mathcal{Q}(t,.)$ converges to $0$ as $m \to +\infty$ from the same arguments giving the continuity of $V_k^{(cs)}$.

%

\subsection{Uniqueness of the Gibbs measure}

The uniqueness of the Gibbs measure associated with a system of transition probabilities $\Probc{\omega(n)}{\sif{n-1}}$ is given by the following theorem \cite{fernandez-maillard:05}.

\begin{theorem}\label{TMaillard}
Let:
$$m(\mathds{P})=\inf_{n \in \mathds{Z}} \inf_{\omega \in \seq{\cA}{-\infty}{n}} \Probc{\omega(n)}{\sif{n-1}}$$
and
$$v(\mathds{P})=\sup_{m' \in \mathds{Z}} \sum_{n \geq m'} var_{n-m'}[\Probc{\omega(n)}{.}].$$
If $m(\mathds{P})>0$ and $v(\mathds{P})< \infty$, then there exist at most one Gibbs measure consistent with it.
\end{theorem} 

We have not been able to prove that those conditions hold in general. We only give here the following argument. Assume that the transition probabilities are differentiable\footnote{Note that Costantini \cite{costantini-etal:06} proved that the first hitting time density of a general diffusion process is differentiable with respect to a time dependent boundary  perturbations, and computed explicitly the derivative.} functions of  $x(n,\omega) \deq \pare{V^{(d)}(n,\omega), \mathcal{Q}(n,\omega))}$, and that the derivative is upper bounded in norm.
Then, the variation of $\Probc{\omega(n)}{\sif{n-1}}$ is bounded by the variation of $x(n,\omega)$ and this variation is exponentially decaying (this follows from the proof of continuity of $V^{(d)}(n,\omega)$ and $\mathcal{Q}(n,\omega)$). As a consequence, $v(\mathds{P}) < \infty$. 

To ensure that $m(\mathds{P})>0$, one needs to slightly modify the model, replacing the usual reset definition where the membrane potential stays
$0$  during the refractory period, by a reset to a random value, as done in  \cite{cessac:11a}.\\

\bibliographystyle{plain}	
\bibliography{biblio,odyssee}	 
\end{document}